\documentstyle[prl,preprint,aps,epsfig]{revtex}
\tightenlines
\begin{document}

\newcommand{\adag}{a^{\dag}}
\newcommand{\atil}{\tilde{a}}
\def\frp#1{${#1\over2}^+$}
\def\frm#1{${#1\over2}^-$}
\def\g{\noindent}

\def\mev{\hbox{\ MeV}}
\def\kev{\hbox{\ keV}}
\def\lambdabar{{\mathchar'26\mkern-9mu\lambda}}
\def\lambdabarrr{{^-\mkern-12mu\lambda}}

\newcommand{\bea}{\begin{eqnarray}}
\newcommand{\eea}{\end{eqnarray}}
\newcommand{\ba}{\begin{array}}
\newcommand{\ea}{\end{array}}
\newcommand{\be}{\begin{equation}}
\newcommand{\ee}{\end{equation}}
\newcommand{\nn}{\nonumber}
\newcommand{\cl}[1]{\begin{center} {#1} \end{center}}
 
\draft
\title{Quantum Statistical Model of Nuclear Multifragmentation \\ 
in the   Canonical Ensemble Method   }

 \author{A.S.~Parvan\dag\S, V.D.~Toneev\dag\ddag
 ~and M.~P{\l}oszajczak\ddag}
\address{\dag\ Bogoliubov Laboratory of Theoretical Physics,  
Joint Institute for Nuclear Research, \\ 141980 Dubna, Russia}
\address{$\S$\ Institute of Applied Physics, Moldova Academy of 
Sciences, MD-2028 Kishineu, Moldova}
\address{\ddag\ Grand Acc\'{e}l\'{e}rateur National d'Ions Lourds (GANIL),
CEA/DSM - CNRS/IN2P3, BP 5027, F-14076 Caen Cedex 05, France }

%\date{\today}

\maketitle

\begin{abstract}
\parbox{14cm}{\rm  A quantum statistical model of nuclear multifragmentation is
proposed. The recurrence equation method used within the canonical ensemble
makes the model solvable and transparent to physical assumptions and 
allows to get  results without 
involving the Monte Carlo technique. The model exhibits the first
order phase transition. Quantum statistics
effects are clearly seen on the microscopic level of occupation numbers but
 are almost washed out for global thermodynamic variables and the averaged 
observables studied.  In the latter case, the recurrence relations for  
multiplicity distributions of both intermediate-mass and all fragments are derived
and the specific changes in the shape of multiplicity distributions in the narrow
region of the transition temperature is stressed. The temperature domain
favorable to  search for the HBT effect is noted.} 
\end{abstract}

\vfill
\newpage

\section{Introduction}
A considerable interest is paid to heavy ion collisions which 
result in the formation of residual nuclei with the excitation energy 
about  $5 - 10 \ AMeV$. In this range of excitation energies 
 one can probe an intriguing possibility of 
 the nuclear liquid-gas phase transition. Below these energies the 
produced system is deexcited by the evaporation of particles, 
mainly nucleons, as would be expected from a hot
liquid drop, whereas at higher excitation energies the excited source explodes
with simultaneous formation of many Intermediate Mass Fragments (IMF). The way
from the production of a residual nucleus to the formation of a multi-fragment
system (the multifragmentation process) is a complex 
phenomenon where a relative role of dynamical and equilibrium effects 
is still not completely understood. Such a situation 
gives rise to the variety of approaches for describing the many-facet nuclear
multifragmentation~: from an evaporative picture~\cite{F90} and simple 
percolation models~\cite{C86,Bauer86}
till a refined  statistical treatment~\cite{RK88,Gross90,Bondorf95,LPT99}  and 
complicated dynamical models~\cite{A91,BP94,FS97,OR97,KD97} 
(see the cited above review-articles  and references quoted therein).

In this work we  develop the Quantum Statistical Model (QSM) of
multifragmentation based on the canonical formalism.   Quantum statistical
treatment has been applied earlier to the multifragmentation process in
Refs.~\cite{S81,HS88,TT94,K94,Gulminelli97}
but within the grand canonical ensemble providing the conservation 
 of proton and neutron numbers in the mean.  The canonical method, which is 
more  suitable for the description of multifragmentation of a 
finite system, corresponds to the exact conservation of baryon number. 
However till now it was used only  for the case of classical 
Boltzmann fragments~\cite{Bondorf95,ChaseC49,ChaseC52}. The
mathematical basis of our consideration is the recursive technique 
\cite{Riordan58} which allows one to eliminate the Monte Carlo
calculations\cite{Bondorf95} and to present results in an analytical form being
rather similar to that for the Boltzmann case \cite{ChaseC49,ChaseC52}. The
model developed in this paper addresses  the issue 
of quantum effects in the multifragmentation process. 
Although these effects {\it a priori} cannot be large, in some
particular cases they may be important.  For example, it was demonstrated that
the yield of $Li$-isotopes, which is used in the studies of the caloric curve
behavior to determine the
system temperature by means of the isotope-ratio method,  is 
sensitive to quantum statistics~\cite{Gulminelli97}. Intensity interferometry
for identical fragments~\cite{HBT}  may give
certain information on the size of fragmenting sources what is of primary  
importance for understanding the multifragmentation process. It is also
well-known that  Bose-Einstein correlations can be the source of salient 
short-ranged phase-space correlations~\cite{BPB95,ZC99}. In general, the study 
of quantum systems of identical particles with incomplete information
may give intriguing and counterintuitive results even in simple cases as
exemplified in~\cite{CC99}.

This paper is divided as follows. Thermodynamics of ideal quantum gas of
identical particles is considered in Sec.~2. In this section we introduce 
the method of recursive relations. 
The canonical ensemble approach is generalized in Sec.~3
 to the case of mixture of different quantum gases for a given composition
of its components. Based on these  results, the QSM is introduced in Sec.~4. 
Technical details of calculations are given in the Appendixes. 
The partition function, various thermodynamic quantities as well as
multiplicity fragment distributions and their moments 
  are derived analytically via the  recurrence relations. Illustrative 
numerical results of the QSM are also presented and discussed in this
section. Finally, concluding remarks are given in Sec.~5.

\section{Perfect gas of identical particles}

Let us consider an ideal gas of $N$ identical particles, either 
bosons or fermions. We want to find the canonical ensemble 
partition function and the ensemble averaging of particle multiplicity.  
Our consideration will be based on constructing the recurrence 
relations which enable an exact calculation of both the statistical 
partition function and the ensemble averaging~\cite{Riordan58}.

Let a system of ideal bosons (fermions) be enclosed in the volume $V$ at the
temperature  $T$. Without any loss of generality we can limit ourselves to
the case of spinless particles, {\it i.e.}, bosons and fermions differ only  by 
the symmetry  of the wave function with respect to permutation of identical
particles.   A microscopic 
state of the system is determined by a set of occupation numbers of
single-particle levels  $\{\nu_{\vec p}\}$, so  in the considered state
  there are $\nu_{\vec p}$ particles with momentum
$\vec p$~\cite{Huang87}.  Single-particle energies are given as
$ \varepsilon_{\vec p} = {\vec p}^2/2 m $
where $\vec p(p^{(x)},p^{(y)},p^{(z)})$  is the eigenvalue 
of the momentum operator of a single particle
\bea \label {a2}
%p^{(a)}=\frac{2 \pi \hbar}{L} n^{(a)}, \; \; \;\;\; \; \;
p^{(a)}= \Delta p^{(a)} \ n^{(a)}~~, \; \; \;\;\; \; \;
n^{(a)}=0,\pm 1,\pm 2,\ldots~~, \; \;\; \; \; \; \;
a=x,y,z \;\; \; \; \; \;\;
\eea
and $\Delta p^{(a)}$ is a  step of the momentum grid towards the direction
$a$~:
\bea \label{5_1}
\Delta p^{(a)} =
\frac{2 \pi \hbar}{L}~~, \;\; \; \; \; \;\;\;\; \; \; \; \;\; L=V^{1/3}~~.
\eea

In the second quantization method, Eq.~(\ref{a2}) determines an infinite set of
discrete values of momentum components of $\vec p$. Every discrete value of
 $\vec p$ corresponds to a single-particle state with 
energy $\varepsilon_{\vec p}$ and
to a particle number $\nu_{\vec p}$. The total particle number
 $N$ and the total energy of this microscopic state
 $E_{\textstyle\{\nu_{\vec p}\} N}$  are
\bea
N &= & \sum_{ \vec p} \ \ \nu_{\vec p}  \label {a4} \\
E_{\textstyle\{\nu_{\vec p}\} N} & = & \sum_{\vec p} \ \  \varepsilon_{\vec
p} \ \nu_{\vec p}~~. \label {a3} 
\eea
The  allowed occupation numbers are
\bea \nu_{\vec p} & = & 0,1,2,\dots,N \; \;
\;\;\; \; \;\; \; \; \mbox{(for bosons)}      \label {a5}   \\ \nu_{\vec p}
& = & 0,1 \;\;\;\;\;\;\;\;\;\;\;\;\;\;\;\;\;\;\;\;\;\;\;\; \mbox{(for
fermions)}~~.             \nonumber
\eea
The canonical ensemble partition function for the perfect Bose (Fermi) gas of 
 $N$ identical particles is~\cite{Huang87}
\bea \label {a6} 
Q_{N}=
\sum_{\textstyle\{\nu_{\vec p}\}, \sum\limits_{\vec p} \nu_{\vec p} = N } \
\ \ e^{\textstyle -\beta E_{\textstyle \{\nu_{\vec p}\}N } }
\eea
where the total energy $E_{\textstyle \{n_{\vec p}\} N}$ is defined by
Eq.~(\ref{a3}), the Lagrange multiplier $\beta (\equiv 1/T)$ and the sum runs 
over all sets of occupation numbers (\ref{a5}), 
 taking into account the conservation of the total particle number $N$.  
The expectation value of an operator $O$ over an ensemble of $N$ 
identical particles can be found in a standard way:
\bea
\label {a7} \langle O \rangle_{N} = \frac{1}{Q_{N}}  \ \sum_{\textstyle
\{\nu_{\vec p}\}, \sum\limits_{\vec p} \nu_{\vec p} = N } \  \ \ O_{
\{\nu_{\vec p}\}} \ e^{ -\textstyle \beta E_{\textstyle \{\nu_{\vec p}\} N}}.
\eea
The partition function (\ref{a6}) can be treated as a generating
function for moments of occupation number distributions. Then the average 
value (\ref{a7}) for different moments of occupation number distributions 
can be found in a following way. Let us introduce a new variable
\bea
x_{\vec p} = e^{ - \textstyle\beta \varepsilon_{\vec p} } 
\label {a17}
\eea
in the partition function  (\ref{a6}) 
\bea Q_{N} = \sum_{\textstyle\{\nu_{\vec p}\}, \sum\limits_{\vec p} \nu_{\vec
p} = N } \ \ \ \prod_{\vec p} \ x_{\vec p}^{\nu_{\vec p}} \ \ .
\label {a16}
\eea
Then, the mean value and highest moments of occupation number distribution 
 can be expressed through the derivatives of the partition function 
(\ref{a16}) with respect to $x_{\vec p}$
\bea
\langle \nu_{\vec p} \rangle_{N} & = & \frac{1}{Q_N} \ x_{\vec p} \ 
\frac{\partial Q_N }{\partial x_{\vec p}} \label {a18}    \\
\langle \nu_{\vec p} \ \nu_{{\vec
p}^\prime}\rangle_{N} & = & \frac{1}{Q_N} \ \ x_{\vec p} \
\frac{\partial}{\partial x_{\vec p}} \ x_{{\vec p}^\prime} \
\frac{\partial}{\partial x_{{\vec p}^\prime}} \ \ Q_N
\label {a19}  \\
\langle \nu_{\vec p} \ \nu_{{\vec p}^\prime} \ 
\nu_{{\vec p}^{\prime\prime}} \rangle_{N} 
& = & \frac{1}{Q_N} \ \ x_{\vec p} \
\frac{\partial}{\partial x_{\vec p}} \ x_{{\vec p}^\prime} \
\frac{\partial}{\partial x_{{\vec p}^\prime}} \ 
\ x_{{\vec p}^{\prime\prime}} \
\frac{\partial}{\partial x_{{\vec p}^{\prime\prime}}} \ 
\ Q_N~~. \label {a19t}
\eea
The frequently used dispersion of the occupation numbers is
\bea
\label {a20} \Delta(\nu_{\vec p} \ \nu_{{\vec p}^\prime})_{N}  = \langle
\nu_{\vec p} \ \nu_{{\vec p}^\prime}\rangle_{N} - \langle \nu_{\vec p}
\rangle_{N} \ \langle \nu_{{\vec p}^\prime} \rangle_{N} \nonumber \\ \qquad
{} = x_{\vec p} \ \frac{\partial}{\partial x_{\vec p}} \ x_{{\vec p}^\prime}
\ \frac{\partial}{\partial x_{{\vec p}^\prime}} \ \ln Q_N~~.
\eea

Let us show now that the partition function $Q_N$  in the occupation
number representation  (\ref{a6}) is reduced to the multinomial 
sum~\cite{Riordan58,Abramowitz65} which can be solved by means of recursive
equations. The comparative analysis of explicit expressions for the 
partition function $Q_N$  and the mean multiplicity
$\langle \nu_{\vec p} \rangle_{N}$ for two systems having  $N$ and $(N-1)$ 
particles reduces to the following relations (see Appendix A)
\bea \label {a8}
\langle \nu_{\vec p} \rangle_{j} \ Q_j \ = \
e^{ -\textstyle \beta \varepsilon_{\vec p}} \  Q_{j-1} \
\left[ 1 \mp   \
\langle \nu_{\vec p} \rangle_{j-1} \right]~~,
\;\;\;\;\;\;\;\;\;\;\;\;    j=1,2,\dots,N 
\eea
with the initial values~: $Q_0=1$ and  $\langle \nu_{\vec p} \rangle_{0}=0$.
 The sign $"+"$/$"-"$ in Eq.~(\ref{a8}) corresponds to the 
 Bose-Einstein/Fermi-Dirac statistics, respectively.
Eq.~(\ref{a8}) is valid for both finite and infinite set of single-particle
levels.  Let us multiply Eq.~(\ref{a8}) by~: $(\mp 1)^{N-j} \ e^{
-\textstyle \beta \varepsilon_{\vec p} (N-j)}$, and sum both left hand side
(l.h.s) and right hand side (r.h.s) of this
equation over $j$ from $1$ to $N$.  This results in the following
representation  for the mean occupation numbers~:
\bea \label {a9}
\langle \nu_{\vec p} \rangle_{N} &=&
\frac{1}{Q_N} \
\sum_{j=1}^{N} \ \
(\mp 1)^{N-j} \ e^{ - \textstyle\beta (N-j+1)\varepsilon_{\vec p}} \
Q_{j-1} \nonumber \\
\label {a10}
&=& \frac{1}{Q_N} \
\sum_{j=1}^{N} \ \
(\mp 1)^{j+1} \ e^{ - \textstyle\beta j \varepsilon_{\vec p} } \
Q_{N-j} \ .
\eea
By the substitution of (\ref{a10}) into the total particle number 
conservation condition 
(\ref{a4}), we arrive at the related set of recurrence equations
 for the partition functions  (\ref{a6}) of systems with different number of
particles~\cite{Riordan58}
\bea
Q_N & = &
\frac{1}{N} \ \sum_{j=1}^{N} \ \
f_j \ Q_{N-j} \ ,
\; \; \; \; \;\; \; \; \; \;\; \;    Q_0=1~~,   \label {a11} 
\eea
where
\bea
f_j & = &
(\mp 1)^{j+1} \
\sum_{\vec p} \ e^{ - \textstyle\beta j \varepsilon_{\vec p} } 
\label {a12}
\eea
This solves the problem of finding the canonical ensemble partition  
function. The sum in (\ref{a12}) runs over the whole infinite set of 
eigenvalues of the momentum operator of a single particle.
Inserting different $Q_j$'s into the recurrence relations (\ref{a11}) 
 gives rise to the multinomial sum
\bea \label {a13} Q_{N} =
\sum_{\textstyle\{n_j \},\sum\limits_{j=1}^{N} \ j n_{j} = N } \  \ \
\prod_{l=1}^{N} \ \ \frac{f_l^{n_l}}{l^{n_l} \ n_l !} 
\eea
which is the well-known quantum-group Meyer expansion for the ideal 
Bose (Fermi) gas of identical particles in the canonical ensemble method 
($n_j$ are integer numbers)~\cite{Huang87}.
One should note that the recurrence relations (\ref{a11}) are satisfied for any
multinomial sum (\ref{a13}), independently of an explicit form of the 
$f_l$ variable~\cite{Riordan58}.  The recurrence relations for the case of the  
ideal gas of $N$ identical
 fermions were obtained recently by Pratt~\cite{Pratt9905055} 
using of somewhat different technique.

The mean occupation number is given by Eq.~(\ref{a10}). To get the second
 moment of the occupation number distributions, let us compare the r.h.s. of
 Eqs.~(\ref{a10}) and (\ref{a18}). Using the definition
 (\ref{a17}) of  $x_{\vec p}$ variable, we find the following relation
between the first derivative of the partition function  $Q_N$
with respect to $x_{\vec p}$ and the partition functions of systems with
smaller number of particles
\bea  \label {a21}
x_{\vec p} \ \frac{\partial Q_N }{\partial
x_{\vec p}} = \sum_{j=1}^{N} \ \ (\mp 1)^{j+1} \ x_{\vec p}^j \
%e^{ -\textstyle\beta j \varepsilon_{\vec p} }
\ Q_{N-j}~~.
\eea
Substituting (\ref{a21}) into
 Eq.~(\ref{a19}) and using again (\ref{a17}), we obtain finally
\bea  \label {a22}
\langle \nu_{\vec p} \nu_{{\vec p}^\prime}\rangle_{N}  =
\frac{1}{Q_N}  \
\sum_{i=1}^{N} \ \sum_{j=1}^{N-i}  \ (\mp 1)^{i+j} \
e^{ - \textstyle\beta i \varepsilon_{\vec p} } \
e^{ - \textstyle\beta j \varepsilon_{{\vec p}^\prime} } \ Q_{N-i-j}
\nonumber          \\
\qquad {} +
\delta_{\vec p {\vec p}^\prime} \ \frac{1}{Q_N} \
\sum_{j=1}^{N} \ \ (\mp 1)^{j+1} \
e^{ - \textstyle\beta j \varepsilon_{\vec p} } \ j \ Q_{N-j}~~.
\eea
The knowledge of the mean occupation numbers (\ref{a10}), 
their second moments  (\ref{a22}) or dispersions (\ref{a20})  
is sufficient for evaluating most of the thermodynamic  quantities.
Higher moments of occupation numbers can be found, if necessary, 
by the same procedure as presented above.  

Similar results can be obtained for the limiting case of an ideal gas of
classical Boltzmann particles in the canonical ensemble. One should take into
account the "correct Boltzmann counting"~\cite{Huang87}, {\it i.e.},  that 
for an arbitrary set of occupations numbers satisfying (\ref{a4}) there are 
generally more ways to construct an ensemble as compared to the Bose gas
because the interchange of any distinguishable particles  leaves the
set $\{\nu_{\vec p}\}$ unchanged  
\bea \label {a23}
Q_{N}^B= \frac{1}{N !} \
\sum_{\textstyle\{\nu_{\vec p}\}, \sum\limits_{\vec p} \nu_{\vec p} = N } \
\ \ \frac{N!}{\prod\limits_{\vec p} \nu_{\vec p}!} \ e^{ - \beta
E_{\textstyle\{\nu_{\vec p}\} N}}
\eea
where the total energy $E_{\textstyle \{\nu_{\vec p}\} N}$ 
is given by Eq.~(\ref{a3}) and the sum runs  over all allowed values of
occupation numbers~: $\nu_{\vec p}=0,1,\dots,N$,  taking into account the
conservation of the total particle number (\ref{a4}).  This sum is the
multinomial Newton sum which can be calculated exactly~\cite{Huang87}
\bea \label {a24}
Q_{N}^B= \frac{1}{N !} \ \
\left(\sum_{\vec p} \
e^{ - \textstyle\beta\varepsilon_{\vec p}}\right)^N .
\eea
The partition function for the ideal Boltzmann gas (\ref{a24}) corresponds to
a single term of the multinomial sum (\ref{a13}) of quantum Bose gas with 
$n_1=N$ and $n_j=0$ for $j \ne 1$
\bea \label {a25}
Q_{N}^{B}= \frac{\omega^N}{N !}~~, \;\;\;\;\;\;\;\;\;\;\;\;
\omega \equiv f_1 =
\sum_{\vec p} \ e^{ - \textstyle\beta\varepsilon_{\vec p}}~~.
\eea
Therefore,  all results presented above for the Bose gas  and expressed
through functions  $f_j$ and $e^{ - \textstyle\beta j \varepsilon_{\vec p}} $,
can be immediately applied to the ideal Boltzmann gas after a simple substitution
\bea \label {a26}
\left\{
\begin{array}{rcl}
e^{ - \textstyle\beta j \varepsilon_{\vec p}} & \to & \delta_{j,1} \
e^{ - \textstyle\beta  \varepsilon_{\vec p}}
\;\;\;\;\;\;\;\;\;\;\;\;  j=1,2,\dots,N   \\
f_j & \to & \delta_{j,1}  \ \omega 
\end{array}
\right.
\eea
where   $\delta_{j,1}$ is the Kronecker  symbol.
Thus,  in the limiting case of the classical Boltzmann statistics only the 
term with $j=1$ survives  in each sum
over $j$ in Eqs. (\ref{a9})--(\ref{a12}), (\ref{a21}), (\ref{a22}).
For example, inserting (\ref{a26}) into Eqs.~(\ref{a11}) results in the 
following recurrence equations for an ideal Boltzmann gas
\bea \label {a28}
Q_{N}^B= \frac{1}{N} \  \omega \ Q_{N-1}^B 
\eea
from where directly follows Eq.~(\ref{a25}). Using (\ref{a26}),
(\ref{a25}) and (\ref{a28}) we have for the mean occupation number 
\bea \label {a29}
\langle
\nu_{\vec p} \rangle_{N}^B = e^{ - \textstyle\beta \varepsilon_{\vec p} }  \
\frac{Q_{N-1}^B}{Q_N^B} \ = \ N \ \frac{
e^{ - \textstyle\beta \varepsilon_{\vec p} } }
{\sum\limits_{\vec p} \ e^{ - \textstyle\beta\varepsilon_{\vec p}}}~~.
\eea

Analogously, the second momentum of the 
occupation number distribution reduces to the form
\bea
\label {a32} \langle
\nu_{\vec p} \nu_{{\vec p}^\prime}\rangle_{N}^B  = N (N-1) \ \frac{ e^{ -
\textstyle\beta \varepsilon_{\vec p} } \ e^{ - \textstyle\beta
\varepsilon_{{\vec p}^\prime} } } {\left(\sum\limits_{\vec p} \ e^{ -
\textstyle\beta\varepsilon_{\vec p}}\right)^2} + \delta_{\vec p {\vec
p}^\prime} \ N \ \frac{ e^{ - \textstyle\beta \varepsilon_{\vec p} } }
{\sum\limits_{\vec p} \ e^{ - \textstyle\beta\varepsilon_{\vec p}}} 
\eea
which is well-known in statistical mechanics~\cite{Huang87}.

Let us now consider fluctuation of particle number distributions 
over different quantum states. The mean squared occupation numbers  
$\langle  \nu_{\vec p}^2 \rangle_{N}$ for Bose and Fermi statistics 
are given in Eq.~(\ref{a22}) which can be rewritten as follows
\bea  \label {fl2}
\langle \nu_{\vec p}^2 \rangle_{N}  =
\frac{1}{Q_N} \
\sum_{j=1}^{N} \ \ (\mp 1)^{j+1} \
e^{ - \textstyle\beta j \varepsilon_{\vec p} } \ Q_{N-j} \
[ \mp (j-1)+j ] \ .
\eea
Using Eqs.~(\ref{a10}) and (\ref{fl2}) we get for the occupation number 
dispersion in the Fermi gas 
\bea \label {fl3}
\langle (\Delta \nu_{\vec p})^2 \rangle_{N} \equiv 
\langle ( \nu_{\vec p} - \langle \nu_{\vec p} \rangle_{N} )^2
\rangle_{N}    =
 \langle  \nu_{\vec p} \rangle_{N} \ \left[ 1 -
\langle  \nu_{\vec p} \rangle_{N} \right ]~~.
\eea
This expression coincides exactly with the appropriate expression 
calculated within the grand
canonical method~\cite{Landau}. This  is due to the equality~:
$ \langle  \nu_{\vec p}^2 \rangle_{N} = \langle  \nu_{\vec p} \rangle_{N}$,
which follows from  (\ref{a10}) and (\ref{fl2}).

Similar consideration for the Bose gas gives
\bea  \label {fl4}
\langle (\Delta \nu_{\vec p})^2 \rangle_{N} =
\frac{1}{Q_N} \
\sum_{j=1}^{N} \ \ (\mp 1)^{j+1} \
e^{ - \textstyle\beta j \varepsilon_{\vec p} } \ (2j-1) \ Q_{N-j} -
\langle \nu_{\vec p} \rangle^2_{N}~~.
\eea
For the case of the Boltzmann gas we have from Eqs.~(\ref{a29}) and (\ref{a32}) 
\bea \label {fl5}
\langle (\Delta \nu_{\vec p})^2 \rangle_{N}^B =
\langle  \nu_{\vec p} \rangle_{N}^B \ \left[ 1 -
\frac{\langle  \nu_{\vec p} \rangle_{N}^B }{N} \right ]~~.
\eea
Here the first term corresponds to the grand canonical result~\cite{Landau}~: 
$ \langle(\Delta \nu_{\vec p})^2\rangle_{GC}  = \langle\nu_{\vec p}\rangle_{GC}$, 
where $\langle\dots\rangle_{GC}$ denotes the grand canonical ensemble 
averaging. As is seen,
in the canonical ensemble treatment an additional negative term arises 
which vanishes in the thermodynamic limit $N \to \infty$.

Let us introduce the quantity $\eta$ as an integral measure of correlations
between the occupation numbers~\cite{Landau}
\bea \label {fl6}
\eta =
\frac{    \sum\limits_{\vec p} \
\left[ \
\langle (\Delta \nu_{\vec p})^2 \rangle_{N} -
\langle  \nu_{\vec p} \rangle_{N} \
\right]    }
{\sum\limits_{\vec p} \ \langle  \nu_{\vec p} \rangle_{N} }~~.
\eea
In application to the Fermi gas, using  Eq.~(\ref{fl3}), we have
\bea \label {fl7}
\eta = - \frac{1}{N} \ \sum\limits_{\vec p} \
\langle  \nu_{\vec p} \rangle_{N}^2~~.
\eea
It is seen that  $\eta$ is negative for all values of  $T,V$ and  $\eta=-1$
at the limiting temperature $T=0$. In the grand canonical ensemble for the
Fermi gas we have~:  $\eta = - (1/\langle N\rangle_{GC} ) \ \sum\limits_{\vec p} 
\langle \nu_{\vec p}\rangle^2_{GC}~.$
For the correlation of identical particles in the Bose gas, substituting 
 (\ref{fl4}) into (\ref{fl6}) and using (\ref{a10}),  we have
\bea  \label {fl8}
\eta =
\frac{1}{N} \ \left \{
\frac{1}{Q_N} \
\sum_{j=1}^{N} \ \ f_j \ 2 (j-1) \ Q_{N-j} -
\sum\limits_{\vec p} \ \langle \nu_{\vec p} \rangle^2_{N}
\right \}~~.
\eea
At $T=0$ one finds~: $\eta=-1$, because~:  $Q_N=1$ and $f_j=1$ 
for any $N$  and $j$~.  In the grand canonical ensemble~:
$\eta = +( 1/\langle N\rangle_{GC})  \sum\limits_{\vec p} \
\langle\nu_{\vec p}\rangle^2_{GC}$~.  Similarly, in the Boltzmann 
limit of the Bose gas with using Eq.~(\ref{fl6}), we have~:
\bea \label {fl9}
\eta^B = - \frac{1}{N^2} \ \sum\limits_{\vec p} \
\langle  \nu_{\vec p} \rangle_{N}^2~~.
\eea
In this case the grand canonical method yields  $\eta=0$. At $T=0$ one 
finds~:  $\eta=-1$, because the mean occupation numbers
(\ref{a29}) are equal to~:
$\langle  \nu_{\vec p} \rangle_{N} = N \delta_{\vec p,0}$~. 

Thus, the occupation number dispersion in the canonical ensemble for 
Bose and Boltzmann statistics differs from those in the grand canonical, 
because the statistical independence of a particle in a state $\vec p$  
is broken due to the requirement of the total particle number 
conservation (\ref{a4}). At the same time,  all three statistics predict 
for the value of the correlation coefficient $\eta \to -1$ when $T\to 0$.

\section{Ideal gas mixture of Bose-Fermi fragments}
$\mbox{}$

Let us consider a system consisting of  $N_1$  individual nucleons, 
$N_2$ clusters (fragments) made of two  nucleons and so on until $N_A$ 
fragments of $A$ nucleons,  contained in the volume $V$ at temperature $T$.  
The composition of this mixture is  characterized by a given set of 
fragment numbers  $ (N_1,N_2, \dots, N_A)$ and satisfies the nucleon 
number conservation condition
\bea \label {1}
\sum_{i=1}^{A} \ i \ N_{i}=A~~.
\eea
 A fragment of
any species $i$ follows either  Bose-Einstein or Fermi-Dirac 
statistics and a number of fragments of a given species $N_i$ is
assumed to be conserved at all permutations. Let us derive the thermodynamic
relations for such a mixture of ideal Bose and Fermi fragments within the
canonical ensemble method. 

A microscopic state of an ideal Bose or Fermi gas subsystem of  $i$-th 
species is defined by a set of occupation numbers $\{\nu_{\vec p i}\}$ for 
single-particle levels of the fragment $i$. This implies that the
microscopic state
in question has $\nu_{\vec p i}$ clusters of mass $i$ with momentum
 $\vec p$~\cite{Huang87}.  The single-particle  energies are~: 
$ \varepsilon_{\vec p i}={\vec p i}^2/2m_i$.

The fragment number $N_i$ of every species $i$ and the total energy of quantum  
gas mixture of a given partition or composition~:
$E_{\textstyle\{\nu_{\vec p i}\} N_1,N_2,\dots ,N_A}$, are defined by the
following expressions (cp. to (\ref{a4}) and (\ref{a3}) )~:
\bea
N_{i} &= & \sum_{ \vec p} \ \ \nu_{\vec p i}~~, \; \; \;\;\; \; \;
i=1,2, \dots, A           \label {7}  \\
E_{\textstyle\{\nu_{\vec p i}\} N_1,N_2,\dots ,N_A} & = & \sum_{i, \vec p} \ \
\varepsilon_{\vec p i} \ \nu_{\vec p i}  \label {6}
\eea
with the allowed values of occupation numbers given in (\ref{a5}).

Generalization of the canonical ensemble partition function (\ref{a6}) to 
the case of an ideal gas mixture of non-identical Bose-Fermi particles with the
conservation of a number of fragments of every species 
(\ref{7}) can be written as~\cite{Huang87}
\bea \label {9}
W_{A}(N_1,\dots,N_A)=
\sum_{\textstyle \{\nu_{\vec p i} \}} \  \ \
\displaystyle e^{ -\textstyle \beta E_{\{\nu_{\vec p i}\}N_1,\dots,N_A}}
\eea
where the sum runs over all partitions of occupation numbers  (\ref{a5}) 
\bea \label {10}
\sum_{\textstyle\{\nu_{\vec p i}\}} \ \equiv \  \
\sum_{\textstyle\{\nu_{\vec p 1}\}\{\nu_{\vec p 2}\}\dots,\{\nu_{\vec p
A}\}} \  \ \prod\limits_{i=1}^A \ \delta\left(\sum_{\vec
p} \ \nu_{\vec p i} - N_i\right)
\eea
 satisfying the conservation laws (\ref{7}). 
Generalization of ensemble averaging (\ref{a7}) to the case of an ideal
gas of Bose-Fermi fragments of different masses is
\bea \label {11}
\langle O \rangle_{N_1,\dots,N_A}=
\frac{1}{W_{A}(N_1,\dots,N_A)} \  \
\sum_{\textstyle\{\nu_{\vec p i}\} } \  \
O(N_1,\dots,N_A) \ \
\displaystyle e^{ -\textstyle \beta E_{\textstyle\{\nu_{\vec p
i}\}N_1,\dots,N_A}}~~.  \eea
Since fragments do not interact with each other,
the total partition function (\ref{9}) is reduced to the product of
 partition functions  $Q(N_i)$ for subsystems of identical fragments $N_i$
\bea \label {13}
W_{A}(N_1,\dots,N_A)=
\prod\limits_{i=1}^A  \ Q(N_i)~~.
\eea

Evaluation of the partition function  $Q(N_i)$ for a subsystem of 
 $N_i$ identical Bose (Fermi) fragments of the mass $i$ 
has been considered in the previous section. Then,
Eqs.~(\ref{a6}), (\ref{a11}) and (\ref{a12}) can be rewritten in the 
new notation as follows
\bea Q(N_i)  &=&  \ \
\sum_{ \{\nu_{\vec p i}\}, \sum\limits_{\vec p} \nu_{\vec p i} = N_i  } \  \
\prod\limits_{\vec p}
\displaystyle e^{ -\textstyle \beta \varepsilon_{\vec p i} \nu_{\vec p i}}
 \label {12}  \\
 & = & \  \frac{1}{N_i} \
\sum_{l=1}^{N_i} \  \
f_{il} \ Q(N_i-l) \ ,
\; \;\; \; \; \; \; \; \;\; \; \; \; \; \; \;\; \;
i=1,2, \dots, A     \label {16} \\
f_{il} & = & (\mp 1)^{l+1} \ \sum_{\vec p} \ \
e^{ -\textstyle \beta l \varepsilon_{\vec p i} } 
\label {15}
\eea
where the signs  $(+)$ and  $(-)$  correspond to the
Bose-Einstein and Fermi-Dirac statistics, respectively.

In the same manner, the mean occupation numbers and their second momenta for a
subsystem of $N_i$ ideal Bose (Fermi) identical fragments of the species $i$ 
can be expressed by the canonical ensemble partition function (\ref{12})
for smaller number of fragments of the same species 
(see (\ref{a10}), (\ref{a22}) )
\bea
\langle \nu_{\vec p i} \rangle_{N_i} &=&
\frac{1}{Q(N_i)} \ \
\sum_{l=1}^{N_i} \ \
(\mp 1)^{l+1} \
\displaystyle e^{ -\textstyle \beta l \varepsilon_{\vec p i} } \
\ Q(N_i-l)   \label {17} \\
 \nonumber \\
\label {18}
\langle \nu_{\vec p i} \ \nu_{{\vec p}^\prime i}\rangle_{N_i} &=&
\frac{1}{Q(N_i)} \ \sum_{l=1}^{N_i} \sum_{k=1}^{N_i-l} \ \ (\mp 1)^{l+k} \
\displaystyle e^{ -\textstyle \beta l \varepsilon_{\vec p i} } \
\displaystyle e^{ -\textstyle \beta k \varepsilon_{{\vec p}^\prime i} } \
Q(N_i-l-k)
\nonumber \\
 & &\qquad {} +
\delta_{\vec p {\vec p}^\prime}    \  \
\frac{1}{Q(N_i)} \ \sum_{l=1}^{N_i} \ \ (\mp 1)^{l+1} \
\displaystyle e^{ -\textstyle \beta l \varepsilon_{\vec p i} } \
l Q(N_i-l)~~.
\eea

By the subsequent substitution of the recurrence relations  (\ref{16})
into Eq.~(\ref{13}), the partition of an ideal quantum gas mixture is
reduced to the generalized quantum-group Meyer expansion  
(cp. to Eq.~(\ref{a13}) )
\bea \label {19}
W_{A}(N_1,\dots,N_A)= \ \prod\limits_{i=1}^{A} \
\sum_{ \{n_j \} , \sum\limits_j j n_j = N_i } \  \
\prod\limits_{l=1}^{N_i} \ \
\frac{(f_{il})^{n_l}}{l^{n_l} \ n_l !}~~.
\eea

Comparing the definitions of  ensemble averaging for identical particles 
 (\ref{a7}) to that for a mixture of Bose-Fermi fragments  (\ref{11}),
we can relate the first and second moments of occupation numbers as well as
their dispersions,  to appropriate quantities for 
subsystems of identical fragments:
\bea
\langle \nu_{\vec p i} \rangle_{N_1,\dots,N_A} & = &
\langle \nu_{\vec p i} \rangle_{N_i}            \label {20}     \\
\langle \nu_{\vec p i} \ \nu_{{\vec p}^\prime j}\rangle_{N_1,\dots,N_A} & = &
\delta_{i j} \ \ \langle \nu_{\vec p i} \ \nu_{{\vec p}^\prime j}\rangle_{N_i} +
\ (1-\delta_{i j}) \ \ \langle \nu_{\vec p i} \rangle_{N_i} \
\langle \nu_{{\vec p}^\prime j} \rangle_{N_j}    \label {21} \\
 \Delta(\nu_{\vec p i} \ \nu_{{\vec p}^\prime j})_{N_1,\dots,N_A} & = &
\langle \nu_{\vec p i} \ \nu_{{\vec p}^\prime j}\rangle_{N_1,\dots,N_A} \ - \
\langle \nu_{\vec p i} \rangle_{N_1,\dots,N_A} \ \
\langle \nu_{{\vec p}^\prime j} \rangle_{N_1,\dots,N_A} \ .    \label {22}
\eea
Therefore, to find the moments of occupation number distributions
(\ref{20})--(\ref{22}) for a mixture of quantum gas with fixed composition of
fragments, it is sufficient to know the occupation number moments for
subsystems of $N_i$ identical fragments. The latter are related via 
Eqs.~(\ref{17}), (\ref{18}) to the subsystem partition function 
 $Q(N_i)$ and can be calculated recursively by means of Eq.~(\ref{16}).

Let us turn now to the calculation of general quantities characterizing
thermodynamic behavior of a system. By definition, the average energy of an
ideal gas mixture of Bose-Fermi non-identical fragments is a partial derivative
of the logarithm of the  partition function (\ref{9}) with respect to
the temperature at a constant volume $V$
\bea \label {23}
E(N_1,\dots,N_A) =
\left . T^2 \ \frac{\partial \ln W_{A}(N_1,\dots,N_A)}{\partial T}
\right |_V =
\langle E_{\textstyle\{N_{\vec p i}\}, N_1,\dots,N_A} \rangle_{N_1,\dots,N_A}
\eea
where the total energy expectation value is calculated  over the ensemble 
 (\ref{11}) and the sum runs over all values of the occupation numbers 
 (\ref{a5}). Because the gas is ideal, the average energy of the system is the
sum of subsystem energies
\bea \label {24}
E(N_1,\dots,N_A) & = & \sum_{i,\vec p} \ \  \varepsilon_{\vec p i} \
\langle \nu_{\vec p i} \rangle_{N_1,\dots,N_A}~~.
\eea
Inserting the mean occupation numbers (\ref{20}), expressed through the partition
functions for subsystems of identical Bose-Fermi fragments (\ref{17}), 
 one obtains
\bea
E(N_1,\dots,N_A) & = &
\sum_{i=1}^A \ \frac{1}{Q(N_i)} \  \sum_{l=1}^{N_i} \ \
\overline{\varepsilon_{il}}  \ \ Q(N_i-l)  \label {25}  \\
\overline{\varepsilon_{il}} & = &
(\mp 1)^{l+1} \
\sum\limits_{\vec p}  \ \ \varepsilon_{\vec p i} \
\displaystyle e^{ -\textstyle \beta l \varepsilon_{\vec p i} }~~. \label {26}
\eea
The summation procedure over the  fragment momentum $\vec p$ in
Eq.~(\ref{26})  is presented in the Appendix B.

By the definition, the pressure in the system is 
\bea \label {27}
P(N_1,\dots,N_A) =
\left . T \ \frac{\partial \ln W_{A}(N_1,\dots,N_A)}{\partial V}
\right |_T = -
\langle \frac{\partial E_{\textstyle\{N_{\vec p i}\}, N_1,N_2,\dots
,N_A}}{\partial V} \rangle_{N_1,\dots,N_A}~~.
\eea
Inserting here  the average energy (\ref{6}), we find
\bea \label {28}
P(N_1,\dots,N_A)  & = & -\sum_{i, \vec p} \
\frac{\partial \varepsilon_{\vec p i}}{\partial V} \
\langle \nu_{\vec p i} \rangle_{N_1,\dots,N_A}~~.
\eea
Making use of Eqs.~(\ref{17}), (\ref{20}), the pressure in a mixed system
 (\ref{28}) can be related to the partition functions for the Bose-Fermi 
subsystems of identical fragments (\ref{16})
\bea
P(N_1,\dots,N_A) & = &
\sum_{i=1}^A  \frac{1}{Q(N_i)} \ \sum_{l=1}^{N_i} \ \
\overline{P_{il}}  \ \ Q(N_i-l)     \label {29}    \\
\overline{P_{il}} & = & -
(\mp 1)^{l+1} \
\sum\limits_{\vec p}  \ \
\displaystyle \frac{\partial \varepsilon_{\vec p i}}{\partial V} \
e^{ -\textstyle \beta l \varepsilon_{\vec p i} }~~.   \label {30}
\eea

In a similar way, one can calculate the specific heat of the system which
is defined as a derivative of the average energy  (\ref{23}) 
with respect to temperature at the constant volume
\bea
C_V(N_1,\dots,N_A)  & =&
\left . \frac{\partial E(N_1,\dots,N_A)}{\partial T}
\right |_V  \nonumber \\ &=&
\frac{\langle E^2_{\textstyle\{\nu_{\vec p i}\}, N_1,\dots,N_A}
\rangle_{N_1,\dots,N_A} \ - \ {\langle E_{\textstyle\{\nu_{\vec p i}\},
N_1,\dots,N_A} \rangle^2_{N_1,\dots,N_A}} }{T^2}~~.
\label {31}
\eea

Inserting (\ref{6})  into  (\ref{11})
and using (\ref{20})--(\ref{22}), we express the specific heat for  
mixture of an ideal quantum gas as 
\bea \label {32}
C_V(N_1,\dots,N_A) & = & {\beta}^2
 \ \sum_{i,j,\vec p,{\vec p}^\prime}
\ \Delta(N_{\vec p i} \nu_{{\vec p}^\prime j})_{N_1,\dots,N_A} \  \
\varepsilon_{\vec p i} \ \varepsilon_{{\vec p}^\prime j}~~.
\eea
Hence, the specific heat can be expressed via the partition functions of identical 
Bose-Fermi fragment subsystems (\ref{12}), if one applies
Eqs.~(\ref{17}), (\ref{18}) and (\ref{20}), (\ref{21})
\bea \label {33}
C_V(N_1,\dots,N_A)  &=& \beta^2
\sum_{i=1}^A  \frac{1}{Q(N_i)} \   \sum_{l,k} \
\overline{\varepsilon_{il}} \
\overline{\varepsilon_{ik}} \ \left[
Q(N_i-l-k)  - \frac{1}{Q(N_i)}   Q(N_i-l)  Q(N_i-k) \right]
\nonumber \\
& & + \ \
\beta^2 \
\sum_{i=1}^A  \frac{1}{Q(N_i)} \  \sum_{l=1}^{N_i} \
l \ \overline{\varepsilon_{il}^2} \  Q(N_i-l)~~.
\eea
In the above equations, the quantity $\overline{\varepsilon_{il}}$ is 
defined by Eq.~(\ref{26}) and the average squared energy of a fragment is
\bea \label {34}
\overline{\varepsilon_{il}^2} & = &
(\mp 1)^{l+1} \
\sum\limits_{\vec p}  \ \ \varepsilon_{\vec p i}^2 \
\displaystyle e^{ -\textstyle \beta l \varepsilon_{\vec p i} }~~.
\eea
Evaluation of average values of $\overline{\varepsilon_{il}}$ and
$\overline{\varepsilon^2_{il}}$ is given in the Appendix B.

Let us consider now the Boltzmann limit of an ideal gas mixture of Bose-Fermi 
fragments. The partition function of the ideal Boltzmann gas 
of non-identical fragments taking into account
the conservation law  for fragment numbers of every species  (\ref{7}) is 
(cp. to (\ref{a24}))~:
\bea \label {c36}
W_{A}^B(N_1,\dots,N_A)=
\frac{1}{\prod\limits_{k=1}^{A} N_k !} \ \
\sum_{\textstyle\{\nu_{\vec p i}\}} \  \
\prod\limits_{k=1}^A  \left(\frac{N_k!}
{\prod\limits_{\vec p} \ \nu_{\vec p k} !}\right)
\displaystyle e^{ -\textstyle \beta E_{\textstyle\{\nu_{\vec p i}\},
N_1,N_2,\dots ,N_A}} 
\eea
where total energy $E_{\textstyle\{\nu_{\vec p i}\}, N_1,N_2,\dots ,N_A}$
is given again by Eq.~(\ref{6}) and occupation numbers take values~:
 $\nu_{\vec p i}=0,1,\dots,N_i$. Similarly to the quantum partition function
 (\ref{9}), the classical partition function (\ref{c36}) can be expressed as a
product of  partition functions
 $Q^B(N_i)$ for subsystems of  $N_i$ identical Boltzmann fragments of mass
 $i$. As was shown in the preceding Section, the partition function 
 $Q^B(N_i)$ for the subsystem of $N_i$ Boltzmann identical fragments of
mass $i$, which is obtained from the appropriate quantum equation
by the simple substitution of (\ref{a26})
\bea \label {c37}
Q^B(N_i)= \frac{\omega_{i}^{N_i}}{N_i!}~~, \;\;\;\;\;\;\;\;\;\;\;\;\;\;\;\;\;
\omega_i \equiv f_{i1} =
\sum_{\vec p} \ e^{ - \textstyle\beta\varepsilon_{\vec p i}} 
\eea
corresponds to a single term of the multinomial sum (\ref{a13}) with 
$n_{i1}=N_i$ and $n_{ij}=0$ for $j\ne1$ $(i=1,2,\dots ,A)$~\cite{Huang87}.
So, the substitution of (\ref{a26}) into the recurrence relations (\ref{16})
leaves only one term   
\bea \label {c38}
Q^B(N_i)= \frac{1}{N_i} \ \omega_{i} \ Q^B(N_i-1)~~, \;\;\;\;\;\;\;\;\;\;\;\;\;\;
i=1,2,\dots,A 
\eea
and the partition function  (\ref{19}) for the mixed Boltzmann gas of fragments
takes very simple form:
\bea \label {cc40}
W_{A}^B(N_1,\dots,N_A)=
\prod\limits_{i=1}^{A}   \
\frac{\omega_{i}^{N_i}}{N_i!}~~.
\eea
Eqs.~(\ref{20})--(\ref{22}) for the moments of the occupation numbers remains
unchanged but the evaluation of the corresponding momentum components is
simplified noticeably (see (\ref{a29}), (\ref{a32})). Taking into account
Eqs.~(\ref{26}) and (\ref{c37}) for the average energy (\ref{25}), we have  
\bea \label {c41}
E^B(N_1,\dots,N_A) =  \sum_{i=1}^A \ \  \overline{\varepsilon_{i}}  \
N_i 
\eea
where
\bea
\overline{\varepsilon_i}  =  \frac
{\sum\limits_{\vec p}  \ \ \varepsilon_{\vec p i} \
e^{ -\textstyle \beta \varepsilon_{\vec p i}}}
{\sum\limits_{\vec p} \ \
e^{ -\textstyle \beta \varepsilon_{\vec p i}}}~~.   \label {c42}
\eea
From Eqs.~(\ref{29}), (\ref{30}) and (\ref{c37}), one finds that the  
pressure in a mixture of the ideal Boltzmann gas is
\bea \label {c43}
P^B(N_1,\dots,N_A) =  \sum_{i=1}^A \ \  \overline{P_i}  \ N_i 
\eea
where
\bea
\overline{P_i} = - \frac
{\sum\limits_{\vec p}  \ \
\displaystyle \frac{\partial \varepsilon_{\vec p i}}{\partial V} \
e^{ -\textstyle \beta \varepsilon_{\vec p i}}}
{\sum\limits_{\vec p} \ \
\displaystyle e^{ -\textstyle \beta \varepsilon_{\vec p i}}}~~. \label {c44}
\eea
Specific heat of the system in the Boltzmann limit is obtained from 
Eqs.~(\ref{33}), (\ref{34}), (\ref{26}) and (\ref{c37}) 
\bea \label {c45}
C_V^B(N_1,\dots,N_A) = \beta^2  \ \sum_{i=1}^A \  \
(\overline{\varepsilon_{i}^2} - {\overline{\varepsilon_i}}^2)
\ N_i  
\eea
with
\bea               
\overline{\varepsilon_i^2} = \frac
{\sum\limits_{\vec p}  \ \ \varepsilon_{\vec p i}^2 \
e^{ -\textstyle \beta \varepsilon_{\vec p i}}}
{\sum\limits_{\vec p} \ \
e^{ -\textstyle \beta \varepsilon_{\vec p i}}}~~. \label {c46}
\eea
These relations will be used below when describing the thermodynamics of 
the quantum statistical model of the multifragmentation.

\section{Quantum statistical model of nuclear multifragmentation}
\subsection{Model formulation}

Based on the recurrence technique discussed above, 
we shall formulate now a quantum  model of fragmentation of a finite 
nuclear system into nucleons and nucleon clusters. This quantum
 fragmentation model is constructed in the the framework of the 
canonical ensemble description in the  occupation number representation. 
Along this way, we shall closely follow the physical approximations used 
in the popular statistical fragmentation models in the canonical 
ensemble~\cite{Bondorf95,Bondorf852,Bondorf853}. In particular, the
partition function of a fragmenting system is defined as a sum over  all
 fragment partitioning, each terms of which is taken with the weight 
associated with the partition function of a given fragment composition.

The  problem of the partitioning of an integer $A$ into a sum of integer
numbers is known in the number theory for a long 
time~\cite{Riordan58,Abramowitz65}.
 Some multifragmentation models~\cite{Mekjian90,LiM90} even employed this 
mathematical result for the description of mass fragment distributions 
assuming equal probability for any fragment composition.  
A partition of $A$ nucleons into clusters
 $(N_1,\dots,N_A)$ corresponds to the fragmentation of a nucleus $A$  into 
 $N_1$  individual nucleons, $N_2$ fragments made of 2 nucleons
  {\it etc.} with the nucleon
conservation constraint (\ref{1}). To construct a multifragmentation model, 
it is necessary to define a probability of any partition
 $(N_1,\dots,N_A)$  with the natural condition that the sum of partial
probabilities over  all final states equals one.
In the statistical multifragmentation models~\cite{Bondorf95}, the
non-normalized probability of a fragment configuration is given by its
partition function. This is just the quantity 
$W_A(N_1,N_2,\dots ,N_A)$ studied in Sec.~3.

Strictly speaking, the ideal gas approximation does not address the problem
 of the nuclear multifragmentation process properly, especially with respect 
to a possible relation of multifragmentation to the liquid-gas phase
transition. The physics content of the model can be however improved by noting
that the decisive condition allowing to develop the general scheme of recurrence
equations is the additivity of the total system energy with respect to subsystem
contributions.   So, the interaction of  $i$ nucleons bound in a single cluster
can be taken into account by adding  the fragment binding energy $b_i$ to the
single particle energy:
\bea
\varepsilon_{\vec p i} & = & \frac{{\vec p}^2}{2 m_i} + b_i~~.
\label {3}
\eea
In principle, some mean-field interaction energy depending only on $i$ can 
be also included into Eq.~(\ref{3}). In particular, such a form has the Coulomb
interaction energy between fragments in the Wigner-Seitz approximation
\bea \label {64}
E^{W-S}_i = \frac{3}{5} \  \frac{A^2 e^2}{4 R} \
\left(\frac{i}{A} \right) \
\left[ 1- {\left(\frac{i}{A}\right)}^{2/3}\right]~~,
\; \; \; \; \;\; \; \; \;\; \; \; \;
R = {\left(\frac{3 V}{4 \pi}\right)}^{1/3} 
\eea
which is used frequently in the statistical description of multifragmentation
 \cite{Bondorf95,Bondorf852,Bondorf853}. 

Repulsive residual nuclear interaction between fragments can be approximately 
accounted for in equilibrium statistical mechanics by the excluded volume 
method~\cite{Huang87,Marmier70}. Since a real fragment occupies some finite
volume  $v_i$, therefore only a reduced 
 volume is reachable for their free motion: 
\bea V_f = V -
\sum_{i=1}^{A} v_i \ N_i~~.  \label{4t}
\eea
It is then assumed that the system may be still treated as 
consisting of idealized point-like fragments in $V_f$. 
It is worth noting that a simple replacement in thermodynamic equations~:  
$V \to V_f$,  needs certain care. Appearance of an additional 
$N_i$ dependence of the partition function  may result in 
the loss of thermodynamic self-consistency. Due to the lack of consistency
 some basic relations of thermodynamics, such as those relating a derivative 
of the thermodynamic potential to the particle number or the pressure,  may 
be violated~\cite{RGSG91,Saeed,Shanenko}. Moreover, as suggested by (\ref{4t}),
the  multiplicity dependence influences  the probability of a given fragment 
partition which is proportional to the (micro)canonical weight of a given 
partition and therefore should be properly accounted for.  Recent discussions 
of this problem can be found in Refs.~\cite{Majumder99,Raduta99,Cole97,GKK99}.
In the QSM, the excluded volume effect is  ignored and $V$ is treated as a 
free variable parameter.

Thus, the canonical ensemble partition function in the QSM of multifragmentation
can be written as follows~:
\bea \label {35}
Z_{A} &=&
\sum_{\{N_j\}, \sum\limits_j j N_j =A} \ \
W_A(N_1,\dots,N_A)                     \nonumber    \\
&=&  \sum_{\{N_j\}, \sum\limits_j j N_j =A} \ \
\sum_{\{\nu_{\vec p i}\}} \  \
e^{ -\displaystyle \beta E_{\textstyle\{\nu_{\vec p i}\} N_1,\dots,N_A}} 
\eea
where the sum runs over all possible fragment partitioning $(N_1,N_2,\dots ,N_A)$
subject to the constraint (\ref{1}). The partition
function for a given fragment configuration is described by Eq.~(\ref{9})
together with  the conservation laws (\ref{7}) for each fragment $i$.  
Because the partition function $W_A(N_1,\dots,N_A)$ 
 can be expressed through the multinomial sum (\ref{19}) depending on
variables (\ref{15}), we have
\bea \label {36_1}
Z_{A}=
\ \sum_{\{N_j\}, \sum\limits_j j N_j =A} \ \
\prod\limits_{i=1}^{A} \
\sum_{\{n_k\}, \sum\limits_k k n_k = N_i } \  \
\prod\limits_{l=1}^{N_i} \ \
\frac{(f_{il})^{n_l}}{l^{n_l} \ n_l !}~~.
\eea
The expectation value of a given operator $O$ over full ensemble of a system
decaying into ideal Bose and Fermi fragments is defined in the QSM as follows 
(cp. to (\ref{11}))~:
\bea \label {37}
\langle O \rangle_{A} &=&
\frac{1}{Z_A}
\ \sum_{\{N_j\}, \sum\limits_j j N_j =A } \ \
\langle O \rangle_{N_1,\dots ,N_A} \ W_A(N_1,\dots ,N_A)
\nonumber \\ &=& \frac{1}{Z_A}
\ \sum_{\{N_j\}, \sum\limits_j j N_j =A } \ \
\sum_{\{\nu_{\vec p i}\}} \  \
O\{\nu_{\vec p i}\} \ \
 e^{ -\textstyle \beta E_{\textstyle\{\nu_{\vec p i}\}N_1,\dots,N_A}}~~.
\eea

Let us find the recurrence relations for $Z_A$. According to  (\ref{37}), the
average occupation numbers can be rewritten in the following form
\bea \label {38}
\langle \nu_{\vec p i} \rangle_{A} = \
\frac{1}{Z_A} \ \
\sum_{\{N_j\}, \sum\limits_j j N_j =A} \ \
\langle \nu_{\vec p i} \rangle_{N_1,\dots,N_A}  \
\prod\limits_{l=1}^{A} \ \ Q(N_l)~~,
\;\;\;\;\;\;\;\;\;\;\;\; i=1,2,\dots,A 
\eea
where Eq.~(\ref{13}) was used for $W_A(N_1,\dots,N_A)$. The insertion  of
 (\ref{20}) and then (\ref{17}) into Eq.~(\ref{38}) and  the exchange of 
the summation order over  $l$ and $\{N_j\}$ followed 
by the extension of the sum over $l$ to the maximal value  $N_i$ 
by means of the step-function $\theta (N_i -l)$, gives
\bea \label {f1}
\langle \nu_{\vec p i} \rangle_{A} =
\frac{1}{Z_A} \ \sum_{l=1}^{[A/i]}  \ (\mp 1)^{l+1} \
e^{ -\textstyle \beta l \varepsilon_{\vec p i} } 
\sum_{\{N_j\}, \sum\limits_j j N_j =A} \
\theta (N_i -l) \ Q(N_i -l) \
\prod\limits_{j \ne i}^{A} \  Q(N_j)~~.
\eea
The step-function removes here those terms in $N_i$ for which $N_i<l$. So, the
mean occupation number is
\bea \label {f2}
\langle \nu_{\vec p i} \rangle_{A} =
\frac{1}{Z_A}  \sum_{l=1}^{[A/i]}  \ (\mp 1)^{l+1} \
e^{ -\textstyle \beta l \varepsilon_{\vec p i} } 
\sum_{\{N_{j \ne i}\}} \ \sum_{N_i=l}^{[A/i]} \
\delta\left(\sum\limits_{j=1}^{A} j N_j -A\right) \ Q(N_i -l) \
\prod\limits_{j \ne i}^{A} \  Q(N_j)~~.
\eea
Changing the summation indices~: $N_i \to N_i +l$, and noting that
 $N_j=0 $ for all  $j>A-i l$ as follows from the Kronecker symbol, we
finally have
\bea \label {39}
\langle \nu_{\vec p i} \rangle_{A} &=&
\frac{1}{Z_A} \ \sum_{l=1}^{[A/i]}  \ (\mp 1)^{l+1} \
e^{ -\textstyle \beta l \varepsilon_{\vec p i} } \
\sum_{\{N_j\},\sum\limits_{j} j N_j =A-i l} \
\prod\limits_{j=1}^{A-i l} \  Q(N_j)          \nonumber  \\
&=& \frac{1}{Z_A} \ \sum_{l=1}^{[A/i]}  \ (\mp 1)^{l+1} \
e^{ -\textstyle \beta l \varepsilon_{\vec p i} }
\ Z_{A-i l} 
\eea
where Eqs.~(\ref{13}), (\ref{35}) were used to get the second equality in
 (\ref{39}). After summing  over momentum $\vec p$, the mean number of 
fragments of mass $i$ in the system of size $A$  can be related to 
the partition functions for  systems of smaller size~:
\bea
\langle N_{i} \rangle_{A}  =
\frac{1}{Z_A}  \ \
\sum_{l=1}^{[A/i]} \ \ f_{il} \ Z_{A-i l}        \label {48}
\eea
where the variables $f_{il}$ were defined in Eq.~(\ref{15}). Note 
that 'il' denotes a single index corresponding to the product
of 'i' by 'l'.

Using Eq.~(\ref{48}) the recurrence equations for the partition 
function $Z_A$ can be easily derived now from the conservation 
of the total nucleon number  (\ref{1})  
\bea \label {40}
Z_A = \frac{1}{A} \ \sum_{i=1}^{A} \
\sum_{l=1}^{[A/i]} \ i \ f_{il} \ Z_{A-i l} 
\eea
where $Z_0=1$ according to (\ref{35}). The partition function for an arbitrary
 $A$ is obtained by an iterative solution of the recurrence equations
 (\ref{40}) starting from $Z_0=1$
\bea \label {f41}
Z_{A}= \ \sum_{\{n_{il}\}, \sum\limits_{i,l} i l n_{il} =A} \
\ \prod\limits_{i=1}^{A} \ \prod\limits_{l=1}^{[A/i]} \ \frac{\displaystyle
(f_{il})^{n_{il}}} {\displaystyle l^{n_{il}} \ n_{il} !}~~.
\eea
The second moment of occupation numbers can be similarly derived from 
Eq.~(\ref{37}) with subsequent using of Eqs.~(\ref{27}) and (\ref{18}) 
\bea \label {46}
\langle \nu_{\vec p i} \ \nu_{{\vec p}^\prime j}\rangle_{A} = \
\frac{1}{Z_A}  \
\sum_{k,l}  \ \
(\mp 1)^{k+l} \
\displaystyle e^{ -\textstyle \beta k \varepsilon_{\vec p i} }
\displaystyle e^{ -\textstyle
\beta l \varepsilon_{{\vec p}^\prime j} }
\  Z_{A-i k-j l}
\nonumber \\
\qquad {} +
\delta_{ij}    \
\delta_{\vec p {\vec p}^\prime}    \  \
\frac{1}{Z_A} \
\sum_{l} \ \
(\mp 1)^{l+1} \
\displaystyle e^{ -\textstyle \beta l \varepsilon_{\vec p i} }
\  l \ Z_{A-i l}~~.
\eea
 Summing over all microscopic states of a fragment gives the second
moment for the fragment number
\bea
\langle N_{i} \ N_{j}\rangle_{A}  =
\delta_{ij}  \  \frac{1}{Z_A} \
\sum_{l=1}^{[A/i]} \ l \ f_{il} \ Z_{A-il} +
\frac{1}{Z_A} \
\sum_{k,l} \  \ f_{ik} \ f_{jl} \ Z_{A-i k-j l}   \label {49}
\eea
where all functions $Z_1,Z_2,\dots,Z_A$ are calculated from Eq.~(\ref{40}) with
$f_{il}$ defined for the system of $A$ nucleons. The knowledge of the first 
(\ref{48}) and the second (\ref{49}) multiplicity moments allows us to find
 the multiplicity dispersion
\bea \label {50}
\Delta(N_{i} \ N_{j})_{A} & = & \langle
N_{i} \ N_j \rangle_{A} \ - \ \langle N_{i} \rangle_{A} \
\langle N_{j} \rangle_{A}~~.
\eea

Similarly to (\ref{fl3}), one can define the dispersion of 
occupation numbers in a given quantum state $\vec p$
\bea \label {ff1}
\langle (\Delta \nu_{\vec p i})^2 \rangle_{A} =
\langle
( \nu_{\vec p i} - \langle \nu_{\vec p i} \rangle_{A} )^2
\rangle_{A}    =
\langle  \nu_{\vec p i}^2 \rangle_{A} -
\langle  \nu_{\vec p i} \rangle_{A}^2~~.
\eea
 The mean squared term $\langle  \nu_{\vec p i}^2 \rangle_{A}$  can be 
obtained for all statistics  from Eq.~(\ref{46}), which reduces to 
\bea  \label {ff2}
\langle \nu_{\vec p i}^2 \rangle_{A}  =
\frac{1}{Z_A} \
\sum_{j=1}^{[A/i]} \ \ (\mp 1)^{j+1} \
e^{ - \textstyle\beta j \varepsilon_{\vec p i} } \ Z_{A-i j} \
[ \mp (j-1)+j ]
\eea
where the mean occupation numbers are defined by Eq.~(\ref{39}). So, the
dispersion (\ref{ff1}) for the case of Fermi gas  fragments is
\bea \label {ff3}
\langle (\Delta \nu_{\vec p i})^2 \rangle_{A} =
\langle  \nu_{\vec p i} \rangle_{A} \ \left[ 1 -
\langle  \nu_{\vec p i} \rangle_{A} \right ]
\eea
and coincides exactly with the result for the identical fermions
(\ref{fl3}). For the gas of Bose fragments, we have correspondingly (cp.
(\ref{fl8}))~:
\bea  \label {ff4}
\langle (\Delta \nu_{\vec p i})^2 \rangle_{A} =
\frac{1}{Z_A} \
\sum_{j=1}^{[A/i]} \
e^{ - \textstyle\beta j \varepsilon_{\vec p i} } \ (2j-1) \ Z_{A-i j} -
\langle \nu_{\vec p i} \rangle_{A}^2~~.
\eea

Note that the calculation of both the partition function
 $Z_A$ and  ensemble averaging is carried out with the help of recurrence
equations (\ref{40}). However, these equations have still other representation
related to the recurrence relations (\ref{a11}) for a system of $N$ identical 
bosons and fermions. By simple algebraic transformations, Eq.~(\ref{40})
can be reduced to the following form~:
\bea Z_A & = &
\frac{1}{A} \ \sum_{i=1}^{A} \ a_i \ Z_{A-i}  \label {41} 
\eea
where
\bea
a_i & = & \sum_{k=1}^i \ \sum_{l=1}^i \ \delta (i-k l) \ k \ f_{kl}~~.
\label {42}
\eea
 It is seen that
Eq.~(\ref{41}) formally coincides with the recurrence relations (\ref{a11}), 
for which the multinomial sum (\ref{a13}) holds. Therefore, 
the partition function $Z_A$ of the canonical ensemble in the QSM of
multifragmentation can be presented as  the multinomial sum (\ref{a13}) 
with the redefined variables  $a_i$  and an evident change~:
 $N \to A$. The recurrence equations (\ref{41}) can be
considered as a set of $A$ linear equations with  $A$ unknown values 
$Z_1,Z_2,\dots,Z_A$~:
\bea \label {43}
j Z_j - \sum_{i=1}^{j-1} \ a_i \ Z_{j-i} = a_j~~,
\; \; \; \; \;\; \; \; \; \;\; \;    j=1,2,\dots,A~~.
\eea
 Solving (\ref{43}) by the Kramer method, we get the solution for
the partition function  (\ref{35})  as a matrix determinant formed by 
 $a_i$ coefficients (\ref{42})
\bea \label {44}
Z_A= \frac{1}{A !}
\left|
\begin{array}{lcccc}
a_{A}   & -a_1 & -a_2 & \ldots & -a_{A-1} \\
a_{A-1} &  A-1 & -a_1 & \ldots & -a_{A-2} \\
a_{A-2} &  0   &  A-2 & \ldots & -a_{A-3} \\
\vdots  & \vdots & \vdots & \ddots & \vdots \\
a_{1}   &  0   &  0   & \ldots &  1       \\
\end{array}
\right|~~.
\eea
The partition function $Q_N$ for the system of $N$ Bose (Fermi) identical
particles (\ref{a11}) can be also presented in the form of matrix determinant
 (\ref{44}) with coefficients~: 
  $a_i=f_i$, and the evident change~: $A \to N$,  $Z_A \to Q_N$. Thus, the
 multinomial sum  (\ref{a13}) can be unambiguously reduced to the matrix
determinant
 (\ref{44}).

The limiting case of the Boltzmann statistics can also be obtained 
in the way presented above. The explicit form of the canonical ensemble 
partition function for multifragmentation of the Boltzmann fragments is
\bea \label {d65}
Z_{A}^B=  \frac{1}{A!}
\ \sum_{\{N_j\}\sum\limits_j j N_j =A} \ 
\frac{A!}{\prod\limits_{k=1}^{A} N_k !} \ 
\sum_{\textstyle\{\nu_{\vec p i}\}} \  
\prod\limits_{k=1}^A  \left(\frac{N_k!}
{\prod\limits_{\vec p} \ \nu_{\vec p k} !}\right)
\displaystyle e^{ -\textstyle \beta E_{\textstyle\{\nu_{\vec p i}\}
N_1,N_2,\dots ,N_A}} 
\eea
where the total energy $E_{\{\nu_{\vec p i}\}N_1,\dots ,N_A}$ is given by 
Eq.~(\ref{6}) and the occupation numbers take values
 $\nu_{\vec p i}=0,1,\dots,N_i$. The ensemble averaging is also defined up to 
 a permutation factor. Making use of the substitution  (\ref{a26})
for Eqs. (\ref{40}), (\ref{f41}), the partition function (\ref{d65}) results in
the multinomial sum for which the following recurrence relations are 
fulfilled~\cite{Riordan58}
\bea
Z_{A}^B  =  \ \sum_{\{N_{i}\},\sum\limits_j j N_{j} =A} \ \
\prod\limits_{i=1}^{A}
\ \frac{\displaystyle \omega_{i}^{N_{i}}}
{\displaystyle  N_{i} !}   =
  \frac{1}{A} \ \sum_{i=1}^{A} \  \
i \ \omega_{i} \ Z_{A-i}^B~~. \label {df65}
\eea
 $\omega_i=f_{i1}$ in (\ref{df65}) denotes  the partition function for 
ideal fragments of the $i$-th species 
(also the integer numbers $n_{i1}$ were denoted as $N_i$).

The recurrence equations of this type were first applied to the
multifragmentation process by Chase and Mekjian~\cite{ChaseC49,ChaseC52}. These
equations were motivated by the Boltzmann treatment of fragments 
with phenomenologically postulated thermodynamic variable  
$\omega_i$ (the so-called tuning parameter). The partition function 
$Z_A^B$ in the form (\ref{44}) remains unchanged but now~: $a_i=i \omega_{i}$, 
for  $i=1,\dots,A$. The Boltzmann substitution (\ref{a26}) 
into expressions for the mean multiplicity of occupation numbers
 (\ref{39}) and fragments numbers (\ref{48}) gives rise to
\bea \label {d67}
\langle \nu_{\vec p i} \rangle_{A}^B & = &
e^{ -\textstyle \beta  \varepsilon_{\vec p i} } \
\frac{Z_{A-i}^B}{Z_A^B}~~,
\;\;\;\;\;\;\;\;\;\;\;\;\;\; i=1,2,\dots,A    \\
\langle N_{i} \rangle_{A}^B & = &
\omega_{i} \ \frac{Z_{A-i}^B}{Z_A^B}~~.    \label {d68}
\eea
For the second moments of the occupation numbers (\ref{46}), 
their dispersion (\ref{ff1}) and the fragment numbers (\ref{49}), 
we  have similarly
\bea \label {d69}
\langle \nu_{\vec p i} \ \nu_{{\vec p} j}\rangle_{A}^B & =  & \
e^{ -\textstyle \beta \varepsilon_{\vec p i} } \
e^{ -\textstyle \beta \varepsilon_{{\vec p}^\prime j} } \
\frac{Z_{A-i-j}^B}{Z_A^B}  \ +
\delta_{ij}    \
\delta_{\vec p {\vec p}^\prime}    \  \
e^{ -\textstyle \beta \varepsilon_{\vec p i} } \
\frac{Z_{A-i}^B}{Z_A^B}           \\
\label {ff5}
\langle (\Delta \nu_{\vec p i})^2 \rangle_{A}^B & = &
\langle  \nu_{\vec p i} \rangle_{A} \ \left[ 1 -
\langle  \nu_{\vec p i} \rangle_{A}  \right ] +
e^{ - \textstyle 2 \beta \varepsilon_{\vec p i} } \
\frac{Z_{A-2 i}^B}{Z_A^B}  \\
\langle N_{i} \ N_{j}\rangle_{A}^B & =  & \
\omega_{i} \omega_{j} \
\frac{Z_{A-i-j}^B}{Z_A^B}  \ +
\delta_{ij}    \
\delta_{\vec p {\vec p}^\prime}    \  \
\omega_{i} \
\frac{Z_{A-i}^B}{Z_A^B}~~.                 \label {d70}
\eea

Let us now illustrate the influence of quantum statistics  on various
characteristics of nuclear fragmentation in terms of
the QSM. Let us consider three versions of the model when {\it all}
produced fragments are treated as either fermions,  bosons or classical
Boltzmann particles. Note that all fragments are assumed to be spinless, {\it
i.e.}, the spin degeneration factor is put equal one. In the results
presented here, the Coulomb interaction potential is taken into account in the
approximation  (\ref{64}) and experimental values of the binding energies $b_i$
are used everywhere unless opposite is said. Internal fragment excitation is
neglected. Volume of the fragmenting nuclear
system $V$ is regarded as a free parameter and its value is given in terms of
the volume of a system  at the normal nuclear density~:  $V/V_0=V \rho_0 /A$.

The momentum dependence of the mean occupation numbers 
$\langle \nu_{\vec p i}\rangle $ is given in Fig.~1 
for different fragments produced in the multifragmentation of system 
with  $A=200$ nucleons. Bose statistics enhances noticeably the yield of
light fragments with low momentum $(p < 150 \ MeV/c)$ but this enhancement
practically vanishes  for IMF's. 
One should note that to get the observable momentum spectra for fragments, 
the given values of $\langle \nu_{\vec p i}\rangle $  should be multiplied 
by the phase factor $p^2$.

Temperature dependence of the mean fragment multiplicity $\langle N_i\rangle$  
for the same system is presented in Fig.~2. Though the  effect of 
quantum statistics slightly increases with temperature $T$, it is seen 
in fact only for nucleons. This is caused by the fact that one needs many 
identical particles  to get a sizable quantum effect but the mean 
multiplicity for fragments is typically small as can be seen 
from Fig.~2. Note that the influence of quantum statistics 
on the mean fragment multiplicity is considerably less than that of 
uncertainty in the choice of the volume $V$ of the fragmenting system. 
The mean multiplicity of the lightest fragments grows gradually 
with increasing violence of the collision, while the IMF's exhibit
clearly the "rise and fall" behavior observed experimentally~\cite{ALADIN}.  
The maximum  in the $T$-dependence of  $\langle N_i\rangle $  of IMF's 
is slightly shifted towards higher temperatures with  increasing size 
of the fragment.

The "rise and fall" behavior noted above is manifested again in the
$T$-dependence of the total mean IMF multiplicity defined as~: 
 $\langle M \rangle=\sum_{i=6}^{40}\langle N_i\rangle$ (see Fig.~3). 
All three model versions give practically 
the same results and demonstrate rather strong dependence of the maximum  
 on the volume of the system. Due to large contribution of nucleons,
the mean multiplicity of all fragments~:  $\langle m\rangle
=\sum_{i=1}^{A}\langle N_i\rangle$, is an increasing function of $T$ with  
abrupt growth above the threshold of IMF production. 
This peculiar structure  of  $\langle m\rangle (T)$-curve is manifested
stronger for  heavier systems, as follows from the same Fig.~3.

 Fluctuations in the region of phase transition behave in a special manner. To 
approach this issue, one can generalize the correlation characteristic 
$\eta$ (\ref{fl6}) to the case of  fragment mixture
\bea \label {ff6}
\eta_i =
\frac{    \sum\limits_{\vec p} \
\left[ \
\langle (\Delta \nu_{\vec p i})^2 \rangle_{A} -
\langle  \nu_{\vec p i} \rangle_{A} \
\right]    }
{\sum\limits_{\vec p} \ \langle  \nu_{\vec p i} \rangle_{A}  }~~.
\eea
Using (\ref{ff3}), we obtain  for the Fermi gas
\bea \label {ff7}
\eta_i = - \frac{1}{\langle  N_i \rangle_{A}} \ \sum\limits_{\vec p} \
\langle  \nu_{\vec p i} \rangle_{A}^2 
\eea
{\it i.e.}, $\eta_i$ is negative for all values of $T,V$ and  $\eta_A=-1$ in the
limit $T\to 0$.  Using (\ref{ff4}), one obtains for the Bose gas  
\bea  \label {ff8}
\eta_i =
\frac{1}{\langle  N_i \rangle_{A}} \ \left \{
\frac{1}{Z_A} \
\sum_{l=1}^{[A/i]} \ \ f_{il} \ 2 (l-1) \ Z_{A-i l} -
\sum\limits_{\vec p} \ \langle \nu_{\vec p i} \rangle_{A}^2
\right \}~~.
\eea
In the Boltzmann limit, this correlation characteristics is  
\bea \label {ff9}
\eta_i^B =  \frac{1}{\langle  N_i \rangle_{A}} \ \left \{
f_{i2} \ \frac{Z_{A-2 i}^B}{Z_A^B} -
\sum\limits_{\vec p} \
\langle  \nu_{\vec p i} \rangle_{A}^2
\right \} 
\eea
and  differs from the Fermi gas case (\ref{ff7}) by a positive correction term.

In Fig.~4 the temperature dependence of $\eta_i$  is presented
 for various fragments. Indeed, there is a striking difference between the cases 
of Bose and Fermi statistics  and this difference 
is maximal at the temperature corresponding to
the  maximal yield of this fragment $i$ going to zero.
 As noted above, $\eta < 0$ for  the Fermi gas. 
Beside the lightest fragment $i=1$, the Boltzmann limit follows very 
closely the Fermi case and the difference between them becomes negligible 
for heaviest IMF's. Because the nucleon component is dominant and 
$\langle N_1\rangle$ is rather large, its $\eta$-behavior looks like that 
for $\langle N_1\rangle$ identical particles with correlations which are
positive, negative or zero for  bosons,  fermions or
Boltzmann particles, respectively. Deviation of $\eta_1$  from zero value 
for the classical case is due to the admixture of other fragments.   

Let us now consider more general characteristics of correlations by summing 
 $\eta_i$ over a certain interval of $i$. In particular, after summing over 
{\it all} $m$ fragments in the numerator and denominator of (\ref{ff6}) we have
\bea \label {ff10}
\eta(m) =
\frac{  \sum\limits_{i=1}^{A}  \sum\limits_{\vec p} \
\langle (\Delta \nu_{\vec p i})^2 \rangle_{A} -
\langle  m \rangle_{A} }
{\langle  m \rangle_{A}  }~~.
\eea
Analogously, the summation over  IMF's yields
\bea \label {ff11}
\eta(M) =
\frac{  \sum\limits_{i=\mu_1}^{\mu_2}  \sum\limits_{\vec p} \
\langle (\Delta \nu_{\vec p i})^2 \rangle_{A} -
\langle  M \rangle_{A} }
{\langle  M \rangle_{A}  } 
\eea
where $\mu_1=6$ and  $\mu_2=40$ were chosen.

As follows from the results presented in Fig.~5, the 
above mentioned properties of the 
occupation number distributions survive in more global characteristics $\eta
(M)$ and $\eta (m)$. Since the contribution of nucleons is dominant, 
therefore $\eta (m) \approx \eta_1$ for temperatures higher than 
about $2-3 \ MeV$. Note that $\eta (m)$ goes down when 
  $T \to 0$. Under this condition,
the system of $A$ identical particles exists as a unique fragment and $\eta_A$
goes to its limiting value $-1$ for all three statistics as discussed above. In
other words, in the limit $T\to 0$  it behaves like a closed-packed
system~\cite{Landau}. It is natural that all correlation effects caused by
quantum statistics are seen more clearly in the denser systems corresponding to
smaller values of $V$. 

 Rather similar correlation characteristics can be defined for the 
multiplicity distribution of a  fragment $i$ 
\bea \label {ff12}
\gamma_i =
 \Delta (N_i^2)_A - \langle  N_i \rangle_{A} 
\eea
where the dispersion $\Delta (N_i^2)_A$ is calculated according to Eq.~(\ref{50}).
Note that as compared to the definition of  $\eta_i$, the factor 
$\langle  N_i \rangle$ is dropped and the quantity $\gamma_i$, which is  
the "shifted" multiplicity dispersion, may be considered
as a shape distribution characteristics.

The calculated $\gamma_i$ are displayed in Fig.~6. 
For the case of Poisson distribution, the
dispersion (\ref{50}) equals the mean multiplicity  (\ref{48}) and 
therefore $ \gamma_i =0$. Such a situation is realised for a dominating nucleon
component  in the Boltzmann gas mixture where 
 $ \gamma_1 \approx 0$.  Visible differences  appear
for quantum  cases. However, because  in the whole mixture of different 
(non-identical) fragments the fraction of specific identical fragments is small,
therefore all three statistics give similar results.  As can be seen 
from Fig.~6 for  $i\ne 1$ and  temperature just above the threshold, 
$\gamma_i(T)$ is positive and  has a spike-like structure 
 showing that the  fragment mass distribution  is getting much wider
at the temperature around  the phase transition. At higher temperature, 
$\gamma_i(T)$ is negative, {\it i.e.}, the distribution is 
sub-Poissonian~\cite{BPB95}  
and approaches asymptotically  the Poisson distribution.          

By suming over $i$ one can define a more global characteristics for IMF's
\bea \label {ff14}
\gamma(M) =  \sum\limits_{i=\mu_1}^{\mu_2}
\Delta (N_i^2)_A - \langle  M \rangle_{A}
\eea
and for all produced fragments
 \bea \label {ff13}
\gamma(m) =
\sum\limits_{i=1}^{A} \ \Delta (N_i^2)_A - \langle  m \rangle_{A} \ .
\eea
As is seen in Fig.~7, the above mentioned features are found in the  
$\gamma (M)$ and $\gamma (m)$ reduced dispersions. The IMF  
distributions near the transition
temperature  have an enlarged  multiplicity width and this width increases
with increasing the freeze-out density of the fragmenting system.

\subsection{Thermodynamic quantities}

Let us now express the global thermodynamic averages characterizing
 the equation of state of nuclear matter through the partition
functions  $Z_A$. The average energy of the  system is
\bea \label {51}
E =
\left . T^2 \ \frac{\partial \ln Z_{A}}{\partial T}
\right |_V =
\langle E_{\textstyle\{\nu_{\vec p i}\}N_1\dots ,N_A} \rangle_{A}~~.
\eea
After the substitution of the average energy (\ref{6}) into expression for the
ensemble averaging (\ref{37}),  we have
\bea
%\label {52}
E  =  \sum_{i, \vec p} \ \  \varepsilon_{\vec p i} \
\langle \nu_{\vec p i} \rangle_{A} =
\frac{1}{Z_A}  \ \sum_{i=1}^A \   \sum_{l=1}^{[A/i]} \  \
\overline{\varepsilon_{il}}   \ Z_{A-i l}~~.    \label {53}
\eea
Here Eq.~(\ref{39}) was used in getting the second equality and
the function $\overline{\varepsilon_{il}}$ was calculated according to
Eq.~(\ref{26}). Making use of Eq.~(\ref{48}), the average energy in the  system
(\ref{53}) can be related to the multiplicity of fragments and to their average
energies
\bea
E & = &
\sum_{i=1}^A \ \overline{\varepsilon_{i}}   \ \
\langle N_{i} \rangle_{A}          \label {54}       \\
\overline{\varepsilon_{i}} & = &
\frac{1}{\overline{f_i}} \
\sum\limits_{l=1}^{[A/i]} \
\overline{\varepsilon_{il}}   \ Z_{A-i l} 
\label {55}                   \\
\overline{f_{i}} & = &
\sum\limits_{j=1}^{[A/i]} \  f_{ij} \ Z_{A-i j}~~.
\label {z55}
\eea
The evaluation of functions  $\overline{\varepsilon_{il}}$ and $f_{il}$ is
considered in the Appendix B. 

The pressure in the multifragmenting system
\bea
%\label {56}
P =
\left . T \ \frac{\partial \ln Z_{A}}{\partial V}
\right |_T = -
\langle \frac{\partial E_{\textstyle\{\nu_{\vec p i}\}N_1\dots
,N_A}}{\partial V} \rangle_{A} =  -\sum_{i, \vec p} \ \frac{\partial
\varepsilon_{\vec p i}}{\partial V} \ \langle \nu_{\vec p i} \rangle_{A}
\label {57}
\eea
can be related to quantities averaged over single-particle momentum, if
Eq.~(\ref{39}) is used for the mean occupation numbers
\bea
P & = &   \
\frac{1}{Z_A}  \ \ \sum_{i=1}^A \   \sum_{l=1}^{[A/i]} \  \
\overline{P_{il}}  \  Z_{A-i l} =
\sum_{i=1}^A \ \overline{P_{i}}   \ \
\langle N_{i} \rangle_{A}      \label {59}  \\
\overline{P_{i}} & = &
\frac{1}{\overline{f_i}} \
\sum\limits_{l=1}^{[A/i]} \  \
\overline{P_{il}}   \ Z_{A-i l}   \label {60}
\eea
with the function  $\overline{P_{il}}$ calculated using Eq.~(\ref{30}).

The heat capacity at the constant volume can be found in the similar way
\bea
%\label {61}
C_V \  = \
\left . \frac{\partial E}{\partial T}
\right |_V   &=&  \
\frac{\langle E^2_{\textstyle\{N_{\vec p i}\}N_1,\dots ,N_A} \rangle_{A}  \ -
\ {\langle E_{\textstyle\{N_{\vec p i}\}N_1,\dots ,N_A} \rangle^2_{A}}
 }{T^2} \nonumber \\
  \label {62}
 &=&  {\beta}^2
 \ \sum_{i, j, \vec p, {\vec p}^\prime}
\ \Delta(\nu_{\vec p i} \nu_{{\vec p}^\prime j})_{A} \  \
\varepsilon_{\vec p i} \ \varepsilon_{{\vec p}^\prime j} \ .
\eea
 Averaging  (\ref{62})  over momentum and using 
Eqs.~(\ref{50}), (\ref{39}) and  (\ref{46}), we finally obtain
\bea \label {63}
C_V & = &
\beta^2  \ \left[
-E^2 +
\frac{1}{Z_A}  \ \ \sum_{i,j} \   \sum_{k,l} \  \
\overline{\varepsilon_{ik}} \  \overline{\varepsilon_{jl}} \
\ Z_{A-i k-j l} +
\frac{1}{Z_A}  \ \ \sum_{i,l}  \ l \
\overline{\varepsilon_{il}^2} \
\ Z_{A-i l} \right]
\eea
where  $\overline{\varepsilon_{il}}$ and $\overline{\varepsilon_{il}^2}$ are 
 defined by Eqs.~(\ref{26}) and  (\ref{34}), respectively.

The classical limit of the derived thermodynamics relations can be found
according to the procedure described above. The average energy in the QSM for the
Boltzmann fragments is obtained from Eqs.~(\ref{54}), (\ref{55}) after the
Boltzmann substitution (\ref{a26})
\bea
E^B & =
& \sum_{i=1}^A \ \overline{\varepsilon_{i}}   \ \ \langle N_{i}
\rangle_{A}^B \label {d71}
\eea
where $\overline{\varepsilon_{i}}$ is defined by (\ref{c42}).  The pressure is
given by Eqs.~(\ref{59}), (\ref{60})~:
\bea \label {d72}
 P^B & = &   \
\sum_{i=1}^A \   \overline{P_i}  \
\langle N_i \rangle_{A}^B
\eea
and  $\overline{P_{i}}$ is defined by (\ref{c44}).  After applying the same 
procedure and using Eqs.~(\ref{50}), (\ref{d68}), (\ref{d70}), (\ref{d71}), 
the heat capacity (\ref{63}) reduces to
\bea \label {d73}
C_V^B & = &
\beta^2  \ \sum_{i j} \
\Delta^B(N_i N_j)_{A} \
\overline{\varepsilon_{i}}  \
\overline{\varepsilon_j} \ + \
\beta^2  \ \sum_{i=1}^A  \
(\overline{\varepsilon_{i}^2} - {\overline{\varepsilon_i}}^2) \
\langle N_i \rangle_{A}^B 
\eea
where  $\overline{\varepsilon_{i}^2}$ is given by Eq.~(\ref{c46}).

An example of thermodynamic characteristics calculated within the QSM of
multifragmentation is presented in Fig.~8. There is a narrow region
of  temperatures where the average energy per nucleon in the  system,
 $E/A$, is growing very  quickly.  This region clearly corresponds 
to a sharp maximum of the heat capacity $C_V$ at a
constant volume indicating the first order phase transition. The finite
width  of $C_V$ reflects the finite size of the system under consideration. 
The position of the  maximum of  $C_V$ depends 
noticeably on the fragmenting volume and is correlated with the position
of  maximum  in IMF multiplicity (see Fig.~2). One should note 
that the quick increase in  $E/A$ as a function of $T$
is unambiguously related to the "plateau" region of the inverse function 
$T(E/A)$ which is called sometimes the caloric curve~\footnote{To be
exact, the caloric curve is the  $T(E^*/A)$ dependence but the excitation 
energy per nucleon $E^*/A$ differs from  $E/A$ by a constant value which 
is equal to the binding energy of the fragmenting nucleus.} 
~\cite{caloric}. As follows from Fig.~8, the plateau region is more
distinct for heavy systems.

The volume dependence of the pressure is shown in Fig.~9. The presented 
 family of isotherms clearly indicates the first order phase
transition in the fragmenting system. Such a behavior in the QSM is
caused solely by the Coulomb interaction of fragments. If the Coulomb
potential (\ref{64}) is  neglected, then the isotherms follow the case of  
classical ideal gas (see r.h.s. in Fig.~9). One should stress 
also that  the considered characteristics of the equation of state 
are insensitive to quantum statistics effects.

\subsection{IMF multiplicity distribution}
$\mbox{}$
In the canonical ensemble, a nuclear system of $A$ nucleons may have $A$
different  species of fragments~: $1\le i \le A$. A number of intermediate 
mass fragments $M$ involving~:   $\mu_1\le i \le \mu_2$, is defined as
\bea \label {i1}
M=\sum_{i \in \mu} N_i
\eea
where  $i$ is taken from the IMF range $\mu=(\mu_1,\dots,\mu_2)$. So, a 
number of IMF's for the system of $A$ nucleons may take the following values
\bea \label {i2}
M=0,1, \dots, \left[ A/\mu_1 \right]~~.
\eea
The probability to find exactly $M$ fragments in the mass range $\mu$ is 
\bea \label {i3}
P_A(M) = \frac{Z_{A}(M)}{Z_{A}}
\eea
where the partition function $Z_A$ is given by Eq.~(\ref{35}) and the new
partition function  $Z_A(M)$, characterizing the probability of a system to be
in a state with  $M$,  is defined as follows
\bea \label {i4}
Z_{A}(M)= \ \sum_{\{N_j\}, \sum\limits_j j N_j =A} \ \
\delta\left(\sum_{i \in \mu} \ N_{i} - M\right) \
\prod\limits_{l=1}^{A} \ Q(N_l)~~.
\eea
Let us derive the recurrence relations that are needed for calculating 
this new partition function  $Z_{A}(M)$ .

 By the direct account of the restriction (\ref{i1}), let us generalize the 
mean occupation numbers of the $A$ nucleon system to the
case of $M$  IMF's
\bea \label {i5}
\langle \nu_{\vec p i} \rangle_{A}(M)  = 
\frac{1}{Z_A(M)}  \
\sum_{\{N_j\} \sum\limits_j j N_j =A } \
\delta\left(\sum_{i \in \mu} \ N_{i} - M\right) \
\langle \nu_{\vec p i} \rangle_{N_1,\dots,N_A}  \
\prod\limits_{l=1}^{A} \  Q(N_l) 
\eea
where $Q(N_l)$ is related to the variable $x_{\vec p i}$ by Eq.~(\ref{a16}).  
By differentiating the logarithm of the partition function (\ref{i4}) with
respect to $x_{\vec p i}$ (see the definition (\ref{a17})), we get for the mean
occupation numbers (\ref{i5})~: 
\bea \label {i7}
\langle \nu_{\vec p i} \rangle_{A}(M)  =
x_{\vec p i} \ \frac{\partial \ln Z_A(M)}
{\partial x_{\vec p i}}~~.
\eea
The product of the partition functions $Q(N_i)$ for  subsystems of identical 
fragments (\ref{12}) enters the definition of the partition function
(\ref{i4}). Each term of this product satisfies the recurrence relation 
(\ref{a18}), so we have
\bea \label {i8}
x_{\vec p i} \ \frac{\partial Q(N_i)}{\partial x_{\vec p i}} =
\sum_{l=1}^{N_i} \
(\mp 1)^{l+1} \ x_{\vec p i}^l \ Q(N_i-l)~~, \;\;\;\;\;\;\;\;\;\;\;\;
i=1,2,\dots,A~~.
\eea
The direct differentiation of the partition function (\ref{i4}) with
respect to $x_{\vec p i}$ gives for all allowed values of $M$
\bea \label {i9}
x_{\vec p i} \ \frac{\partial Z_A(M)}{\partial x_{\vec p i}} =
\sum_{\{N_j\}, \sum\limits_j j \ N_j =A} \ 
\delta\left(\sum_{k \in \mu} \ N_{k} - M\right) \
\left(x_{\vec p i} \ \frac{\partial \ln Q(N_i)}{\partial x_{\vec p i}}
\right) \
\prod\limits_{l=1}^{A} \ Q(N_l)~~.
\eea
 Substituting  (\ref{i8}) into (\ref{i9}) and  using the method employed above 
(see Eqs.~(\ref{f1})--(\ref{39}) ) to calculate $\langle\nu_{\vec p i}\rangle$, 
one finds for the case $M=0$
\bea \label {i10}
x_{\vec p i} \ \frac{\partial Z_A(M=0)}{\partial x_{\vec p i}} & = &
\sum_{l=1}^{[A/i]} \
(\mp 1)^{l+1} \ x_{\vec p i}^l \ Z_{A-i l}(M=0)~~, 
\;\;\;\;\;\;\;\;\;\;\;\;\;     i \notin \mu       \\
x_{\vec p i} \ \frac{\partial Z_A(M=0)}{\partial x_{\vec p i}} & = & 0~~, 
\;\;\;\;\;\;\;\;\;\;\;\;\;\;\;\;\;\;\;\;\;\;\;\;\;\;\;\;\;\;\;\;\;\;\;\;\;\;\;
\;\;\;\;\;\;\;\;\;\;\;\;\;\;\;\;\;\;\;     i \in \mu~~.
\label {i11}
\eea
By differentiating  (\ref{i9}), we obtain in the case of $M \ne 0$ and  
$i \in \mu$ 
\bea \label {i12}
x_{\vec p i} \ \frac{\partial Z_A(M)}{\partial x_{\vec p i}} & = &
\sum_{l=1}^{\min([A/i],M)} \ \
(\mp 1)^{l+1} \ x_{\vec p i}^l \ Z_{A-i l}(M-l)~~.
\eea
The mean number of fragments $i$ for the $A$ nucleon system with the fixed number
$M$ of IMF's is
\bea \label {i13}
\langle N_{i} \rangle_{A}(M)  =
\sum_{\vec p} \ \
\langle \nu_{\vec p i} \rangle_{A}(M) 
\eea
where the mean occupation numbers are determined by (\ref{i5}), (\ref{i7}). The 
mean fragment number (\ref{i13})  satisfies the restrictions 
 (\ref{1}) and (\ref{i1}). By the substitution of Eqs.~(\ref{i10}), (\ref{i11})
into (\ref{i7}) and then into Eq.~(\ref{i13}) we have for the mean fragment
multiplicity in the case of $M=0$ 
\bea
\langle N_{i} \rangle_{A}(M=0) & = &
\frac{1}{\textstyle Z_A(M=0)} \ \
\sum_{l=1}^{[A/i]} \ \ f_{il} \ Z_{A-il}(M=0)~~, 
\;\;\;\;\;\;\;\;\;\;\;\;  i \notin \mu \label {i14} \\
\langle N_{i} \rangle_{A}(M=0) & = & 0~~,
\;\;\;\;\;\;\;\;\;\;\;\;\;\;\;\;\;\;\;\;\;\;\;\;\;\;\;\;\;\;\;\;\;\;
\;\;\;\;\;\;\;\;\;\;\;\;\;\;\;\;\;\;\;\;\;\;\;\;\;\;\;\;\;\;\;\;
i \in \mu~~. 
\label {i15} 
\eea
Here $f_{il}$ is calculated according to (\ref{15}).  

To find the average fragment multiplicity in the case of $M \ne 0$ 
and $i \in \mu$,  one inserts Eq.~(\ref{i12}) into (\ref{i7}) 
what  after using  (\ref{i13}) results in the following relation~:
\bea \label {i16}
\langle N_{i} \rangle_{A}(M)  =
\frac{1}{Z_A(M)} \ \
\sum_{l=1}^{\min([A/i],M)} \ \
f_{il} \ Z_{A-i l}(M-l)~~.
\eea

In accordance with (\ref{i4}), we have the following relations. For 
systems with the nucleon number $j < \mu_1$ and  the IMF number $M=0$  
\bea \label {i17}
Z_{j}(M=0) = Z_{j}~~,
\;\;\;\;\;\;\;\;\;\;\;\;\;\; j=0,1,\dots,\mu_1-1~~.
\eea
For system with $j < \mu_1\cdot M$ and  $M \ne 0$ one has
\bea \label {i18}
Z_{j}(M) = 0~~,
\;\;\;\;\;\;\;\;\;\;\;\;\;\;  j=0,1,\dots,\mu_1\cdot M-1~~.
\eea
Finally, for the "conditional" partition function $Z_{j}(M)$ in the case of
$M=0$, after the substitution of the mean fragment numbers  (\ref{i14}), 
(\ref{i16}) in the the conservation law of the total number of nucleons
(\ref{1}), we have the following recurrence equations
\bea \label {i19}
Z_{j}(M=0)= \ \frac{1}{j}  \
\sum_{i \notin \mu} \ \sum_{l=1}^{[j/i]} \ i \ f_{il} \ Z_{j-i l}(M=0)~~,
\;\;\;\;\;\;\;\;\;\;\;\;\;\; j=\mu_1,\dots,A~~.
\eea
The recurrence equations for the case of any $M \ne 0$, 
$ (M=1,2,\dots,[j/\mu_1])$ are obtained similarly by making use of 
the restriction (\ref{i1}) and Eq.~(\ref{i16})~: 
\bea \label {i20}
Z_{j}(M)= \ \frac{1}{M} \
\sum_{i \in \mu} \ \sum_{l=1}^{\min([j/i],M)} \ f_{il} \ Z_{j-i l}(M-l) \ ,
\;\;\;\;\;\;\;\;\;   j=\mu_1\cdot M,\dots,A~~.
\eea
If one now divides Eqs.~(\ref{i17}) and (\ref{i20}) by the partition function
 $Z_j$ and applies the probability definition 
(\ref{i3}), we get the recurrence relations for probability to find exactly $M$
fragments among  IMF's. If $M=0$
\bea
P_{j}(M=0) & = & 1~~,  \;\;\;\;\;\;\;\;\;\;\;\;\;\;\;\;\;   
\;\;\;\;\;\;\;\;\;\;\;\;\;\;\;\;\;\;\;\;\;\;\;\; 
\;\;\;\;\;\;\;\;\;\;\;\;\;\;\;\;\;\;
j=0,1,\dots,\mu_1-1                       \label {i21}  \\
P_{j}(M=0) & = &  \frac{1}{j} \ \sum_{i \notin \mu}
\sum_{l=1}^{[j/i]} \ P_{j-i l}(M=0) \ i \ f_{il} \
\frac{Z_{j-i l}}{Z_j}~~, 
 \;\;\;\;\;\;\;\;    j=\mu_1,\dots,A~~.  \label {i23}
\eea
If $M \ne 0$ then
\bea
P_{j}(M) & = & 0~~,  \;\;\;\;\;\;\;\;\;\;\;\;\;\;\;\;\;\;\;\; 
\;\;\;\;\;\;\;\;\;\;\;\;\;\;\;\;\;\;\;\;\;\;\;\;\;\;\;\; 
\;\;\;\;\;\;\;\;\;\;\;\;\;\;\;\;\;\;\;\;
j=0,1,\dots,\mu_1\cdot M-1                   \label {i22}      \\
P_{j}(M) & = &  \frac{1}{M} \
\sum_{i \in \mu}  \sum_{l=1}^{\min([j/i],M)} \
P_{j-i l}(M-l) \ f_{il} \
\frac{Z_{j-i l}}{Z_j}~~,
\;\;\;\;\;\;\;\;\;  j=\mu_1\cdot M,\dots,A~~.          \label {i24}
\eea
In the Boltzmann limit, the IMF multiplicity distributions and related
equations   can be easily derived from the above derived quantum results 
by means of the substitution (\ref{a26}). In particular, the
equalities  (\ref{i17}), (\ref{i18}) for the partition function are not 
changed and Eqs.~(\ref{i19}) and (\ref{i20}) reduce to the following 
equations for $M=0$ and $M\ne 0$, respectively~:
\bea \label {e75}
Z_{j}^{B}(M=0)&=& \ \frac{1}{j}  \
\sum_{k \notin \mu} \  k \ \omega_{k} \ Z_{j-k}^{B}(M=0)~~,
\;\;\;\;\;\;\;\;\;\;\;\;\;   j=\mu_1,\dots,A  \\
 \label {e76}
Z_{j}^{B}(M)&=& \ \frac{1}{M} \
\sum_{k \in \mu} \  \omega_{k} \ Z_{j-k}^{B}(M-1)~~,
\;\;\;\;\;\;\;\;\;\;\;\;\;\;  j=\mu_1\cdot M,\dots,A~~.
\eea
Similarly, the recurrence equations for the probability to find $M$ IMF's 
in the classical model of multifragmentation result from 
Eqs.~(\ref{i21})--(\ref{i24}). If $M=0$
\bea
P_{j}(M=0) & = & 1~~,  \;\;\;\;\;\;\;\;\;\;\;\;\;\;\;\; 
\;\;\;\;\;\;\;\;\;\;\;\;\;\;\;\;\;\;\;\;\;\;\;\;\;\;\;\;\; 
\;\;\;\;\;\;\;\;\;\;\;\; j=0,1,\dots,\mu_1-1  \label{e77}  \\
P_{j}(M=0) & = & \ \frac{1}{j} \ \sum_{k \notin \mu} \
P_{j-k}(M=0) \ k \ \omega_{k} \  \frac{Z_{j-k}}{Z_j}~~,
\;\;\;\;\;\;\;\;\;\;    j=\mu_1,\dots,A~~.  \label {e79}  
\eea
If $M \ne 0$
\bea
P_{j}(M) & = & 0~~,  \;\;\;\;\;\;\;\;\;\;\;\;\;\;\;\;\;   
\;\;\;\;\;\;\;\;\;\;\;\;\;\;\;\;\;\;\;\;\;\;\;\;\;\;
\;\;\;\;\;\;\;\;\;\;\;\;  j=0,1,\dots,\mu_1\cdot M-1 \label{e78}\\
P_{j}(M) & = & \ \frac{1}{M} \ \sum_{k \in \mu} \
P_{j-k}(M-1) \ \omega_{k} \ \frac{Z_{j-k}}{Z_j}~~,
\;\;\;\;\;\;\;\;\;         j=\mu_1\cdot M,\dots,A~~.   \label {e80}
\eea
One should note that the multiplicity distributions of IMF's within the Boltzmann
statistics have been recently studied in Refs.~\cite{Pratt9903006,Pratt9903007}. 
Their equation coincides with our Eq.~(\ref{e80}) 
but Eqs.~(\ref{e77}), (\ref{e79}) for $M=0$ have not been found and
the normalization condition $\sum_M P(M)=1$ were used  to get the full
distribution in~\cite{Pratt9903006,Pratt9903007}.

The multiplicity distributions of IMF's ($\mu_1=6, \ \mu_2=40$) calculated
within the QSM of multifragmentation are presented in Fig.~10. It is of great
interest that the $M$-distribution sharply changes its shape in the
vicinity of the phase transition temperature  defined by the maximum
 of the heat capacity. These results are in full agreement with the
discussion of the reduced dispersions in Figs.~6, 7 and reveal the different 
IMF distributions in different phases as it would be expected for the phase 
transition.   Details of the multiplicity distributions depend
on the mass number $A$ and the size of the system, being weakly sensitive to
quantum statistics of fragments. Some local irregularities and the appearance 
of two maxima in the transition curve of IMF multiplicity distributions 
are caused by the use of experimental values for the fragment binding 
energy.     If the Coulomb fragment interaction is neglected and
a smoothed  approximation is used for the binding energies, the QSM with the
Boltzmann statistics reproduces exactly the numerical results 
in~\cite{Pratt9903006,Pratt9903007}.

\subsection{Total multiplicity distribution}
$\mbox{}$
The total fragment multiplicity $m$ associated with the multifragmentation of
 a system of the size $A$ is defined similarly to  (\ref{i1})
\bea \label {m1}
m=\sum_{i \in \mu}  N_i 
\eea
except that the region $\mu$ is now~:
\bea \label {m2}
\mu=(1,2, \dots, A) 
\eea
{\it i.e.}, the events with $m=0$ are absent due to the nucleon 
number conservation. The probability to find  $m$ fragments is defined 
by the same Eq.~(\ref{i3}) with the conditional probability (\ref{i4}). 
From this follows immediately that in the case of $m=0$ 
the only non-zero partition function is that for $j=0$~:
\bea \label {m17}  
Z_{j}(m) = \delta_{j,0}~~,
\;\;\;\;\;\;\;\;\;\;\;\;\;\;  %%m = 0,
\;\;\;\;\;\;\;\;\;\;\;\;\;\;  j=0,1,\dots,A~~.
\eea
For the partition functions with $j<m$, the following relations are valid
\bea \label {m18}
Z_{j}(m) = 0~~,
\;\;\;\;\;\;\;\;\;\;\;\;\;\;  m=1,2,\dots,A,
\;\;\;\;\;\;\;\;\;\;\;\;\;\;  j=0,1,\dots,m-1~~.
\eea
In this case, the recurrence equations  (\ref{i8})  and the derivative of the
partition function  (\ref{i9}) still hold but the region  $\mu$ is defined by
Eq.~(\ref{m2}). Similar transformation of Eq.~(\ref{i9}) with using (\ref{i8}) 
gives for the particular value $m=1$
\bea
x_{\vec p i} \ \frac{\partial Z_A(m)}{\partial x_{\vec p i}} & = & 0~~,
\;\;\;\;\;\;\;\;\;\;\;\;\;\;\;\;\;\;\;\;\;\;\;\;\;\;\;\;\;\;\;\;\;\;
    i=1,2,\dots,A-1  \label {m10} \\
x_{\vec p i} \ \frac{\partial Z_A(m)}{\partial x_{\vec p i}} & = &
x_{\vec p i} \ Z_{A-i}(m-1)~~,
\;\;\;\;\;\;\;\;\;\;\;    i=A~~.       \label {m11}
\eea
For derivatives (\ref{i9}) with $m=2,3,\dots,A$ we have 
\bea
x_{\vec p i}  \frac{\partial Z_A(m)}{\partial x_{\vec p i}} & = &
\sum_{l=1}^{\min([A/i],m)}  
(\mp 1)^{l+1} \ x_{\vec p i}^l \ Z_{A-i l}(m-l)~~,
\;\;\;\;    i=1,2,\dots,A-m+1   \label {m12} \\
x_{\vec p i}  \frac{\partial Z_A(m)}{\partial x_{\vec p i}} & = & 0~~,
\;\;\;\;\;\;\;\;\;\;\;\;\;\;\;\;\;\;\;\;\;\;\;\;\;\;\;\;\;\;\;\;\;\;\;\;\;\;
\;\;\;\;\;\;\;\;\;\;\;\;\;\;\;\;\;\;   i=A-m+2,\dots,A~~.  
\label {m12_1}
\eea
According to the definition, the mean fragment number of the $A$ nucleon
system with the total multiplicity $m$, for which two restrictions (\ref{1}) 
and (\ref{m1}) are fulfilled, is
\bea \label {m13}
\langle N_{i} \rangle_{A}(m)  =
\sum_{\vec p} \ \
\langle N_{\vec p i} \rangle_{A}(m)~~. 
\eea
 By substituting (\ref{m10}) and (\ref{m11})  into Eq.~(\ref{i7}),  we get
 for the mean fragment number (\ref{m13}) in the case of $m=1$~:
\bea
\langle N_{i} \rangle_{A}(m) & = & 0~~,
\;\;\;\;\;\;\;\;\;\;\;\;\;\;\;\;\;\;\;\;\;\;\;\;\;\;\;\;\;\;\;\;\;\;\;
  i=1,2,\dots,A-1 \label {m14} \\
\langle N_{i} \rangle_{A}(m) & = &
f_{i1} \ \frac{Z_{A-i}(m-1)}{Z_A(m)}~~,
\;\;\;\;\;\;\;\;\;\;\;\;  i=A \label {m15}
\eea
with $f_{i1}$ calculated according to (\ref{15}). This quantity for the
case of $m=2,3,\dots,A$ is obtained by combining
(\ref{m12}), (\ref{m12_1}) and (\ref{i7}) 
\bea \label {m16}
\langle N_{i} \rangle_{A}(m)  & = &
\frac{1}{Z_A(m)} \ 
\sum_{l=1}^{\min([A/i],m)} \ 
f_{il} \ Z_{A-il}(m-l)~~,
\;\;\;\;\;\; i=1,2,\dots,A-m+1 \\
\langle N_{i} \rangle_{A}(m)  & = &  0~~,
\;\;\;\;\;\;\;\;\;\;\;\;\;\;\;\;\;\;\;\;\;\;\;\;\;\;\;\;\;\;\;\;\;\;\;\; 
\;\;\;\;\;\;\;\;\;\;\;\;\;\;\;\;\;\;\;\;\;\; 
i=A-m+2,\dots,A  \label {m16_1}
\eea                                                                    
with $f_{il}$ defined again by (\ref{15}).

As usually, the recurrence relations are derived from the restriction 
 (\ref{1}) to the  $j$ nucleon system. From Eqs.~(\ref{m14}), 
(\ref{m15}) we have for  $m=1$
\bea \label {m19}
Z_{j}(m)= \ \ f_{j1}~~,
\;\;\;\;\;\;\;\;\;\;\;\;\;\; j=1,\dots,A
\eea
and similarly for $m=2,3,\dots,A$, by using Eqs.~(\ref{m16}) and 
(\ref{m16_1}), we have 
\bea \label {m20}
Z_{j}(m)= \ \frac{1}{j} \
\sum_{i=1}^{j-m+1} \ \sum_{l=1}^{\min([j/i],m)} \
i \ f_{il} \ Z_{j-i l}(m-l) \ ,
\;\;\;\;\;\;\;\;\;   j=m,\dots,A~~.
\eea
Eqs. (\ref{m17}), (\ref{m18}) and (\ref{m19}), (\ref{m20}) form a complete
set which is sufficient for calculating the partition function  $Z_A(m)$ and 
therefore the multiplicity distributions $P(m)$.

The limiting case of the Boltzmann statistics  for the total multiplicity
distributions can be obtained from the appropriate quantum equations by means
of the  substitution (\ref{a26}).  For the system of classical fragments,
Eqs.~(\ref{m17}), (\ref{m18}), (\ref{m19}) are not changed and Eq.~(\ref{a26})
results in the following recurrence relations in the case of  
$m=2,3,\dots,A$~:
\bea \label {m21}
Z_{j}^{B}(m)= \ \frac{1}{j}  \
\sum_{i=1}^{j-m+1} \  i \ \omega_{i} \ Z_{j-i}^{B}(m-1)~~,
\; \; \; \; \;\; \; \; \;   j=m,\dots,A~~.
\eea

The total multiplicity distributions of fragments within the QSM are 
exemplified in Fig.~11. General behavior of $m$-distributions is close 
to those noted before while discussing the IMF distributions 
(cp. to Fig.~10), however all curves became more smooth. 
The reason of this behavior is a dominance of  individual nucleons
  what results also in somewhat stronger influence of the quantum
statistics on the multiplicity distributions.

 \section{Conclusions}

The model proposed in this work realizes for the first time the 
quantum statistical approach to the
multifragmentation of excited finite nuclei in the framework of the canonical 
ensemble method. The mathematical basis of the model is the recurrence equation 
technique which allows, within the usually accepted physics assumptions,  
to solve a cumbersome problem of fragment partitioning {\it exactly} 
without involving  complicated and time-consuming Monte Carlo methods.    
The QSM opens a possibility for calculating in the same framework various
microscopic characteristics of occupation numbers, global thermodynamic variables 
specifying the equation of state, and different observables allowing 
for a comparison with the experiment. 
The exactly solvable technique makes such calculations transparent 
to the physics assumptions used. The QSM  includes the limiting case
of the classical Boltzmann statistics and in this case the QSM reduces 
to a class of solvable models
developed earlier~\cite{ChaseC49,ChaseC52,Mekjian90,LiM90}.

In principle, effects of quantum statistics in the QSM are seen on 
the microscopic level 
in quantities like $\langle \nu_i\rangle , \langle \nu_i^2\rangle $, 
especially at temperatures corresponding to the maximal fragment yield.
 The strength of these effects is proportional to a number of available 
identical clusters of a given size. Thus, the effect is  strongest for
nucleons and  is noticeably weaker for the IMF's. It is of
interest that  large uncertainties in the choice of the freeze-out volume
do not kill  these quantum statistical effects in the correlation
characteristics $\eta_i$.

 Global thermodynamic variables, which are insensitive to quantum statistics,
  exhibit nevertheless a rather sharp phase
transition of the first order in the temperature region $\approx 6-8 \ MeV$
depending on the size of the system and on the value of the freeze-out volume 
$V/V_0$. In the QSM, the observed Van der Waals behavior is caused by the
competition between  attractive forces, included in the fragment binding
energies, and  the Coulomb interaction energy. It is noteworthy 
that at not so high temperatures, the value of $V/V_0=3$,
being the standard value for the SMM  calculations of the primary fragment
partition~\cite{Bondorf95}, turns out to be in the thermodynamically 
unstable region~: $\partial P / \partial V < 0$.  
This implies that before  getting an equilibrium solution in the model 
analysis, the Maxwell construction should be applied 
to the isotherms in this unstable region. The use of 
these Maxwellian isotherms is important for extracting true values 
of the freeze-out temperature of a system as well as for
detailed studies of the caloric curve.
  
Fragment multiplicity is a very  interesting observable, even 
though the quantum statistical
effects are practically absent in the averaged quantities. 
The mean number $\langle N_i\rangle $ of IMF's clearly demonstrates 
the "rise and fall" behavior as a function of temperature. Fragment
multiplicity dispersion has a maximum near the transition temperature
corresponding to the maximum of heat capacity $C_V$ as a sign of the
phase transition in finite systems to be dependent on the system size, 
freeze-out volume and  fragment binding energies. 
These features are remarkably well seen in the change of the shape
of the IMF multiplicity distribution in  the transition region. 
   The measurement  of the HBT effect for  IMF's which would be crucial 
for determining the size of a fragmenting system, seems to be a hard
task due to a comparatively small number of identical particles. 
However, the model studies in this paper show where this effect is 
expected to be maximal. In addition, when searching 
for the HBT effect a more detailed quantity, the correlated particle
spectrum $\langle N_i(\vec p) \ N_i(\vec p^{\prime})\rangle$ rather than 
the mean square multiplicity $\langle N_i^2\rangle$  should be  
investigated~\cite{BPB95}. In any case, this issue deserves a special 
discussion.  Making use of protons instead of the IMF's unfortunately  
does not help much because in a real experiment 
this component may be strongly enriched by non-equilibrium particles.

As a final remark, the model developed in this paper takes 
into account the mass number conservation while the conservation of the 
charge number may be quite important for the yield of specific isotopes 
or even for the equation of state. In particular, it was shown that the
liquid-gas transition in the asymmetric nuclear matter is of the second 
order rather than of the first order as one would expect 
for neutron-proton symmetric matter~\cite{MS95}. This process is of
particular interest because experimental data for $Au+C$ at 1 $GeV$ per
nucleon~\cite{Hauger96} shows a much smoother phase transition than the 
first observation in the $Au+Au$ collision at 600 $MeV$~\cite{ALADIN}.  
The recurrence equation technique allows one to consider
simultaneously both mass and charge number conservation. Such a generalization
has been done in Refs.\cite{ChM94,PGT99} but only for the mean multiplicity of
individual fragments within the Boltzmann statistics. Similar extension for
calculating  the multiplicity distributions of the identified IMF's would open a
new way to study critical exponents, intermittence phenomenon and so on. 
This work is in progress now.

\vskip \baselineskip
 
{\bf Acknowledgements~:}
We are thankful to K.K.~Gudima and A.A.~Shanenko  for useful discussions. 
The work was supported by  the CNRS-JINR agreement No 96-28.
\vfill

\appendix{}
\section{A useful recurrence relation}
%$$$$$$$$$$$$$$$$$$$$$$$$$$$$$$$$$$$$$$$$$$$$$$$$$$$$$$$$$$$$$$$$$$$$$$$$

As an auxiliary step, let us define the partition function
$Q_N^{n_{\vec p}}$ for a system of $N$ Bose (Fermi) identical particles
with the fixed number
$n_{\vec p}$ of particles in a given single particle state  $\vec p$
\bea  \label {ab1}
Q_{N}^{n_{\vec p}}  =
\sum_{\{\nu_{\vec p}\}} \
\delta \left(\sum\limits_{\vec p^{\prime}} \nu_{\vec p^{\prime}}-N \right) \
\delta \left(n_{\vec p} - \nu_{\vec p} \right) \
\prod_{\vec p^{\prime}} \
x_{\vec p^{\prime}}^{\nu_{\vec p^{\prime}}} 
\eea
where the conservation laws are accounted for by the Kronecker symbol and 
the allowed occupation numbers $\nu_{\vec p}$  are given by
  (\ref{a5}). Then, the total partition function
(\ref{a16}) for $N$ identical particles  can be expressed via auxiliary one
 (\ref{ab1}) as follows
\bea  \label {ab2}
Q_{N}  =
\sum_{n_{\vec p}=0}^{\{1\},\{N\}} \ Q_{N}^{n_{\vec p}}
\eea
where the upper limit in the sum is equal to $N$ or $1$ for bosons or fermions,
respectively. The sum in  (\ref{ab1}) with respect to  $\nu_{\vec p}$ 
is taken by means of delta functions
\bea  \label {ab3}
Q_{N}^{n_{\vec p}}  &=&
\sum_{\{\nu_{\vec p^{\prime}}\}_{\vec p^{\prime} \ne \vec p}} \
\delta \left(\sum\limits_{\vec p^{\prime} \ne \vec p}
\nu_{\vec p^{\prime}} + n_{\vec p} - N \right) \
x_{\vec p}^{n_{\vec p}} \
\prod_{\vec p^{\prime} \ne \vec p} \
x_{\vec p^{\prime}}^{\nu_{\vec p^{\prime}}}   \nonumber  \\
&=& x_{\vec p}^{n_{\vec p}} \
\sum_{\{\nu_{\vec p^{\prime}}\}} \
\delta \left(\sum\limits_{\vec p^{\prime}}
\nu_{\vec p^{\prime}} -(N - n_{\vec p}) \right) \
\delta \left(\nu_{\vec p} \right) \
\prod_{\vec p^{\prime}} \
x_{\vec p^{\prime}}^{\nu_{\vec p^{\prime}}}~~.
\eea
Comparing the r.h.s of  (\ref{ab3}) to the definition (\ref{ab1}), one
can see that here one can select out the partition function for  
$(N-n_{\vec p})$ particles in the state  $\vec p =0$ 
\bea  \label {ab4}
Q_{N}^{n_{\vec p}}  =
x_{\vec p}^{n_{\vec p}} \ Q_{N-n_{\vec p}}^{n_{\vec p}=0}~~. 
\eea
This will result in the recurrence relations for the partition function 
(\ref{ab1}) which are valid for both  Bose and Fermi particles. Let us 
start with the boson case. After the substitution of  (\ref{ab4}) into
(\ref{ab2}), we have for the total partition function
\bea  \label {ab5}
Q_N =
Q_{N}^{n_{\vec p}=0}  +
\sum_{n_{\vec p}=1}^{N} \
x_{\vec p}^{n_{\vec p}} \ Q_{N-n_{\vec p}}^{n_{\vec p}=0}~~.
\eea
Changing the summation indices  in the second term on the r.h.s. of
 (\ref{ab5}) and using the recurrence relations (\ref{ab4}), we obtain
\bea  
%\label {ab6}
Q_N =
Q_{N}^{n_{\vec p}=0}  +
x_{\vec p} \
\sum_{n_{\vec p}=0}^{N-1} \ Q_{N-1}^{n_{\vec p}} 
 = Q_{N}^{n_{\vec p}=0}  + x_{\vec p} \ Q_{N-1}  
\label {ab7} \eea
where the partition function for $(N-1)$ particles  appears explicitly in
the second term. After differentiating (\ref{ab7}) with respect to $x_{\vec p}$
and taking into account that this derivative of the partition function 
 $Q_{N}^{n_{\vec p}=0}$ equals zero, we arrive at the following differential
recurrence equations:
\bea  \label {ab8}
x_{\vec p} \ \frac{\partial Q_N}{\partial x_{\vec p}} \ = \
x_{\vec p} \ Q_{N-1} + x_{\vec p}^2 \
\frac{\partial Q_{N-1}}{\partial x_{\vec p}}~~.
\eea
The mean occupation numbers can be expressed through the derivative of
the logarithm of the partition function $Q_{N}$ with respect to
 $x_{\vec p}$  (see Eq.~(\ref{a18})). So,  Eq.~(\ref{ab8}) for Bose particles
 reduces to the recurrence equations 
\bea \label {ab9}
\langle n_{\vec p} \rangle_{N} \ Q_N \ = \
x_{\vec p} \  Q_{N-1} \
\left[ 1  +
\langle n_{\vec p} \rangle_{N-1} \right] 
\eea
where  $Q_0=1$ and $\langle n_{\vec p} \rangle_{0}=0$ according to
Eqs.~(\ref{a6}) and (\ref{a7}), respectively.

Let us proceed now to the system of  fermions.
Plugging  (\ref{ab4}) in (\ref{ab2}), we have for the partition functions of 
two  systems with $N$ and $(N-1)$ particles
\bea
Q_N & = &  Q_{N}^{n_{\vec p}=0}  +
x_{\vec p} \ Q_{N-1}^{n_{\vec p}=0}   \label {ab10} \\
\nonumber \\
Q_{N-1} & = &  Q_{N-1}^{n_{\vec p}=0}  +
x_{\vec p} \ Q_{N-1-1}^{n_{\vec p}=0}~~.  \label {ab11}
\eea

To get the differential equations for the system of fermions, let us 
differentiate Eqs.~(\ref{ab10}) and (\ref{ab11}) with respect to  
$x_{\vec p}$, multiply the first equation by  $x_{\vec p}$ and 
second one by $x_{\vec p}^2$ and sum these two~:
\bea  \label {ab12}
x_{\vec p} \ \frac{\partial Q_N}{\partial x_{\vec p}} \ = \
x_{\vec p} \ Q_{N-1} - x_{\vec p}^2 \
\frac{\partial Q_{N-1}}{\partial x_{\vec p}}~~.
\eea
Here we took into consideration that derivatives of $Q_{N}^{n_{\vec p}=0}$ and 
$Q_{N-1}^{n_{\vec p}=0}$ with respect to $x_{\vec p}$ equal zero.  Combining   
(\ref{a18}) and (\ref{ab12}), we obtain the final recurrence equations for the
Fermi system~:
\bea \label {ab13}
\langle n_{\vec p} \rangle_{N} \ Q_N \ = \
x_{\vec p} \  Q_{N-1} \
\left[ 1  -
\langle n_{\vec p} \rangle_{N-1} \right]~~.
\eea
The same initial condition as in the boson case are used here.
One should note that Eqs. (\ref{ab9}) and (\ref{ab13}) for Bose and Fermi
statistics differ only by the sign in front of the second term.

%$$$$$$$$$$$$$$$$$$$$$$$$$$$$$$$$$$$$$$$$$$$$$$$$$$$$$$$$$$$$$$$$$$$$$$$$

\section{Momentum resumation}

Using (\ref{3}), let us present the sums in
(\ref{15}), (\ref{26}), (\ref{30}) and (\ref{34}) in more convenient form 
\bea
f_{il} & = & (\mp 1)^{l+1} \
e^{ -\beta l E_i} \
\sum_{\vec p} \
e^{ -\textstyle \frac{\beta l {\vec p}^2 }{2 m_i } } 
\label {b3}      \\
\overline{\varepsilon_{il}} & = &
(\mp 1)^{l+1} \
e^{ -\beta l E_i} \
\sum\limits_{\vec p}   \ \frac{{\vec p}^2 }{2 m_i }  \
e^{ -\textstyle \frac{\beta l {\vec p}^2 }{2 m_i } } \ + \
E_i \ f_{il} 
\label {b4}    \\
\overline{\varepsilon_{il}^2} & = &
(\mp 1)^{l+1} \
e^{ -\beta l E_i} \
\sum\limits_{\vec p}   \ {\left(\frac{{\vec p}^2 }{2 m_i }\right)}^2  \
e^{ -\textstyle \frac{\beta l {\vec p}^2 }{2 m_i } } \ +  \
2 E_i \ \overline{\varepsilon_{il}}  \ - \ E_i^2 \ f_{il}
\label {b5}   \\
\overline{P_{il}}  & = &
\frac{2}{3} \ \frac{\overline{\varepsilon_{il}} }{V} \ - \
f_{il} \left(\frac{\partial E_i}{\partial V} \ + \
\frac{2}{3} \  \frac{E_i}{V} \right)~~.
\label {b6}
\eea
The summation over  momentum in the above Eqs.~(\ref{b3})--(\ref{b6}) can be
written as 
\bea
\sum_{\vec p} \
e^{ -\textstyle \frac{\beta l {\vec p}^2 }{2 m_i } } & = &
\sum_{\vec n} \
e^{ -\textstyle \frac{\beta l {(\Delta p)}^2 }{2 m_i } {\vec n}^2}
\label {b7}      \\
\sum\limits_{\vec p}   \ \frac{{\vec p}^2 }{2 m_i }  \
e^{ -\textstyle \frac{\beta l {\vec p}^2 }{2 m_i } } & = &
\frac{{(\Delta p)}^2 }{2 m_i }  \
\sum\limits_{\vec n}   \    {\vec n}^2
e^{ -\textstyle \frac{\beta l {(\Delta p)}^2 }{2 m_i } {\vec n}^2} 
\label {b8}    \\
\sum\limits_{\vec p}   \ {\left(\frac{{\vec p}^2 }{2 m_i }\right)}^2  \
e^{ -\textstyle \frac{\beta l {\vec p}^2 }{2 m_i } } & = &
\left(\frac{{(\Delta p)}^2 }{2 m_i }\right)^2  \
\sum\limits_{\vec n}   \    {\vec n}^4  \
e^{ -\textstyle \frac{\beta l {(\Delta p)}^2 }{2 m_i } {\vec n}^2}~~.
\label {b9}
\eea
where $\vec n \equiv (n^{(x)},n^{(y)},v^{(z)})$ and each $n^{(a)}$ 
 runs over all integer numbers (see Eq.~(\ref{a2})).
It is worthwhile to proceed to the momentum integration in  
(\ref{b7})--(\ref{b9}),
though this is  not valid in the whole region  of temperature 
 $T$  and volume $V$. The passage to the integration over momentum
\bea \label {b10}
\sum_{\vec p} \dots = \int \frac{d^3 \vec p}{(\Delta p)^3} \dots 
\eea
is possible under the condition
\bea \label {b11}
\frac{(\Delta p)^2 l}{2 m_i T} \le 1~~.
\eea
In the opposite case, when
\bea \label {b12}
\frac{(\Delta p)^2 l}{2 m_i T} > 1
\eea
this passage can be erroneous and the sums (\ref{b7})--(\ref{b9}) should be
calculated numerically.
If the condition (\ref{b11}) is satisfied, we have for
integrals of (\ref{b7})--(\ref{b9})~: 
\bea
\sum_{\vec p} \
e^{ -\textstyle \frac{\beta l {\vec p}^2 }{2 m_i } } & = &
\frac{1}{l^{3/2}} \ \frac{V}{\lambda_i^3} 
\label {b13}      \\
\sum\limits_{\vec p}   \ \frac{{\vec p}^2 }{2 m_i }  \
e^{ -\textstyle \frac{\beta l {\vec p}^2 }{2 m_i } } & = &
\frac{3}{2} \ T \
\frac{1}{l^{5/2}} \ \frac{V}{\lambda_i^3} 
\label {b14}    \\
\sum\limits_{\vec p}   \ {\left(\frac{{\vec p}^2 }{2 m_i }\right)}^2  \
e^{ -\textstyle \frac{\beta l {\vec p}^2 }{2 m_i } } & = &
\frac{15}{4} \ T^2 \
\frac{1}{l^{7/2}} \ \frac{V}{\lambda_i^3}
\label {b15}
\eea
where the thermodynamic wavelength for a fragment made from
 $i$ nucleons  is
\bea \label {b16}
\lambda_i = \left(\frac{2 \pi \hbar^2}{m_i T}\right)^{1/2}~~,
\;\;\;\;\;\;\;\;\;\;\;\;\;\;\;\;\;\;\;\;\;\;\;\; m_i=i m_N
\eea
and  $m_N$ stands here for the nucleon mass. After the substitution of
 (\ref{b13})--(\ref{b15}) into (\ref{b3})--(\ref{b5}), we obtain
\bea
f_{il} & = & (\mp 1)^{l+1} \
e^{ -\beta l E_i} \
\frac{1}{l^{3/2}} \ \frac{V}{\lambda_i^3} 
\label {b17}      \\
\overline{\varepsilon_{il}} & = &
\left(\frac{3}{2} \ T \ l^{-1} \ + \ E_i \right) \ f_{il} 
\label {b18}    \\
\overline{\varepsilon_{il}^2} & = &
\left(\frac{15}{4} \ T^2 \ l^{-2} \ + \ 3 \ E_i \ T \ l^{-1} \ +E_i^2
\right) \ f_{il}~~.
\label {b19}
\eea

For calculating (\ref{b7})--(\ref{b9}) in the case (\ref{b12}),
let us present these sums in the form more suitable for computing.
Let us introduce the function depending on the integer vector $\vec n$ 
\bea \label {b20}
z_{il}(\sigma) =
\sum\limits_{ n=-\infty}^{\infty}   \    { n}^{\sigma}  \
e^{ -\textstyle \frac{\beta l {(\Delta p)}^2 }{2 m_i } {n}^2}~~,
\; \; \; \; \; \; \; \; \; \; \; \; \; \; \;   \sigma = 0,2,4~~.
\eea
In this notation, the sums (\ref{b7})--(\ref{b9})  are reduced to the following
expressions~:
\bea
\sum_{\vec p} \
e^{ -\textstyle \frac{\beta l {\vec p}^2 }{2 m_i } } & = & \
z_{il}^3(0) 
\label {b21}      \\
\sum\limits_{\vec p}   \ \frac{{\vec p}^2 }{2 m_i }  \
e^{ -\textstyle \frac{\beta l {\vec p}^2 }{2 m_i } } & = &  \
3 \ z_{il}(2) \ z_{il}^2(0) \ \ \frac{(\Delta p)^2 }{2 m_i } 
\label {b22}    \\
\sum\limits_{\vec p}   \ {\left(\frac{{\vec p}^2 }{2 m_i }\right)}^2  \
e^{ -\textstyle \frac{\beta l {\vec p}^2 }{2 m_i } } & = & \
3 \ z_{il}(0) \ \left[ z_{il}(4) \ z_{il}(0) + 2 \ z_{il}^2(2) \right] \ \
\frac{(\Delta p)^4 }{4 m_i^2 }~~.
\label {b23}
\eea
To calculate them, it is sufficient to know functions 
  $z_{il}(0), z_{il}(2)$ and $z_{il}(4)$ in the limited range of  values $n$,  
because exponents go fast to zero with increasing  the absolute value of  $n$.

\vfill
\newpage
\begin{center}
{\bf Figure captures}
\end{center}

Fig.~1. The mean occupation numbers $\langle\nu_{\vec p i}\rangle$ 
of fragments of mass $i$  for   $A=200$ system at 
temperatures of 10 and 15 $MeV$. The dashed, dotted and continuous curves  
correspond to the QSM result obtained under the assumption 
that all fragments are bosons, fermions or Boltzmann particles,
respectively. The system volume is $V/V_0=3$ and $\Delta p = 90$ MeV/c. \\

Fig.~2. The temperature dependence of the mean multiplicity 
$\langle N_{i}\rangle$ for  fragment of mass $i$ in
 the fragmenting system of  size $A=200$. 
For further details, see  the caption of Fig.~1.
The  results are obtained for the two values of the system volume. \\

Fig.~3.The temperature  dependence of the total mean multiplicity
 $\langle m\rangle $ and IMF multiplicity $\langle M\rangle$ for systems with
 $A=100$ and 200. For further details see  the caption of  Fig.~2. \\

Fig.~4. The temperature dependence of the occupation number correlation $\eta_i$ 
of the fragment species  $i$  for the system with $A=200$ at two different 
values of parameter $V/V_0$. For further details see  the caption of Fig.~2.  \\  

Fig.~5. The temperature dependence of the occupation-number correlation parameter  
summed over all fragments $\eta (m)$ and IMF's $\eta (M)$  for   $A=200$. 
  For further details see  the caption of Fig.~4. \\

Fig.~6. The temperature dependence of the shifted multiplicity 
dispersion $\gamma_i$  of the fragment species  $i$  
for the system with $A=200$ at two values of the freeze-out volume 
parameter $V/V_0$. For further details see  the caption of Fig.~2.  \\  

Fig.~7. The temperature dependence of the  shifted multiplicity dispersion 
summed over all fragments $\gamma (m)$ and IMF's $\gamma (M)$  for the  
system with $A=200$. For further details see  the caption of Fig.~4. \\

Fig.~8.  The heat capacity and the total energy vs. temperature for systems with 
$A=200$ (solid line) and 100 (dashed) nucleons 
at two different values of the volume parameter $V/V_0$. 
Dotted curves are calculated for  $A=200$ system neglecting the Coulomb
interaction energy.  The  results for Bose,  Fermi and Boltzmann statistics  
coincide with each other with the accuracy comparable to the line thickness. \\

Fig.~9. The pressure as a function of the volume at the constant temperature $T$ 
for the  systems size $A=100$ and 200  with (l.h.s.) and without (r.h.s.) taking
into account the Coulomb interaction energy. The  results for Bose,  Fermi and 
Boltzmann statistics  coincide with each other with the accuracy comparable 
to the line thickness. \\               

Fig.~10. The IMF multiplicity distributions for systems with $A= 200$ and
 100 nucleons at the given temperature $T$ for two different  
values of the volume parameter~:  $V/V_0 = 6$ (lower panels)  
and $V/V_0 = 3$ (upper panels). Open squares, open circles and filled 
triangles are the QSM result obtained under the assumption that all 
fragments are bosons, fermions or Boltzmann particles, respectively. \\

Fig.~11. The total fragment multiplicity distributions. For further details see  
the caption of Fig.~10. \\

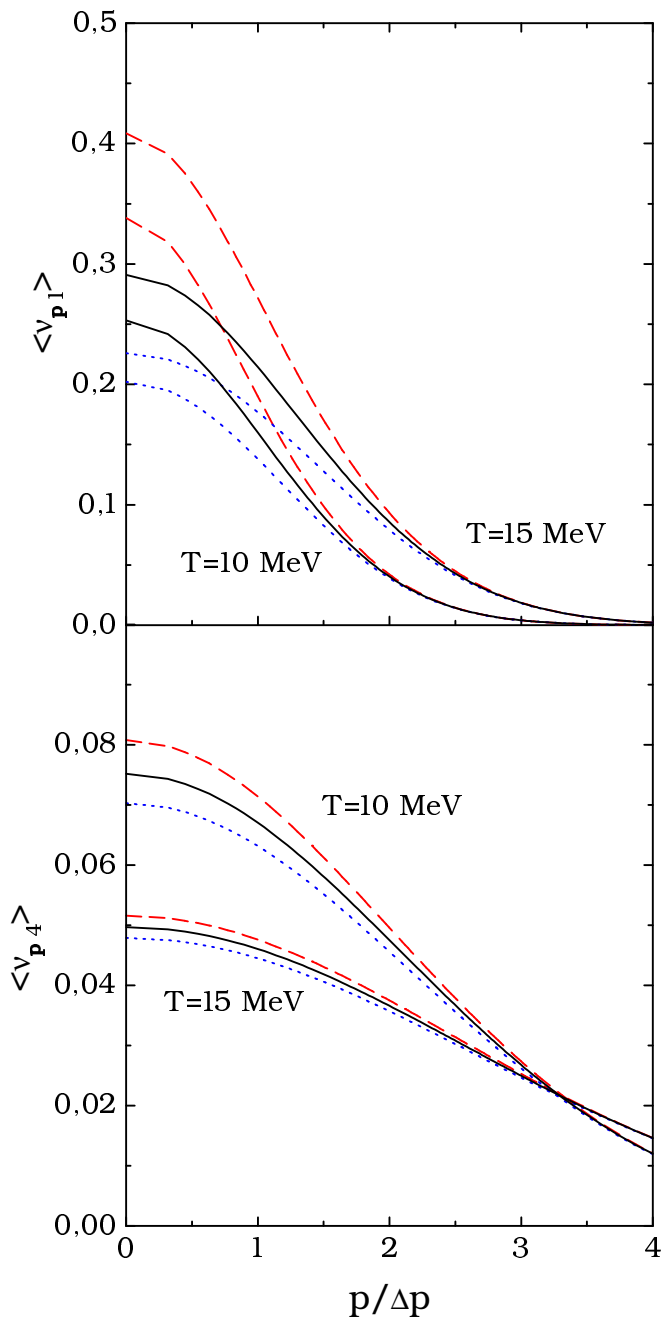
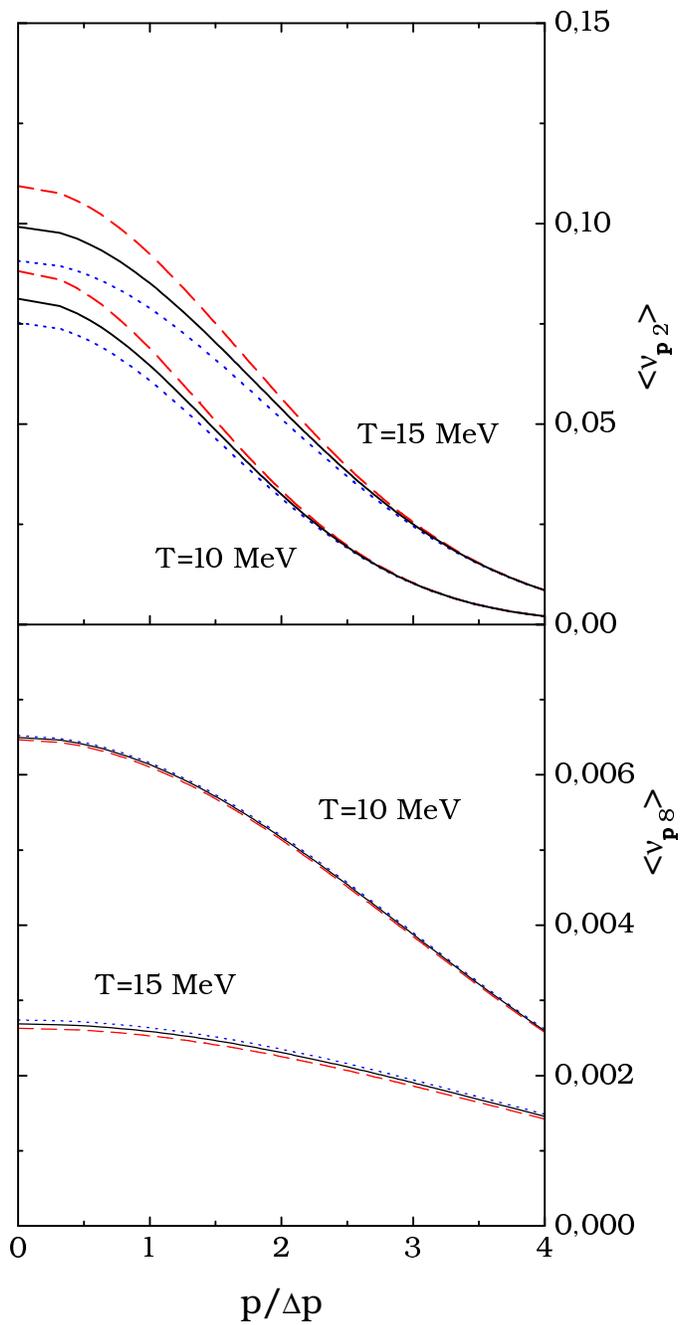
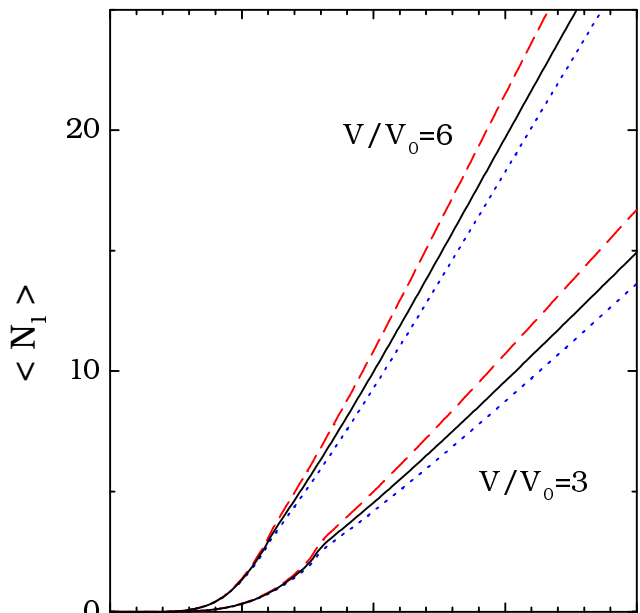
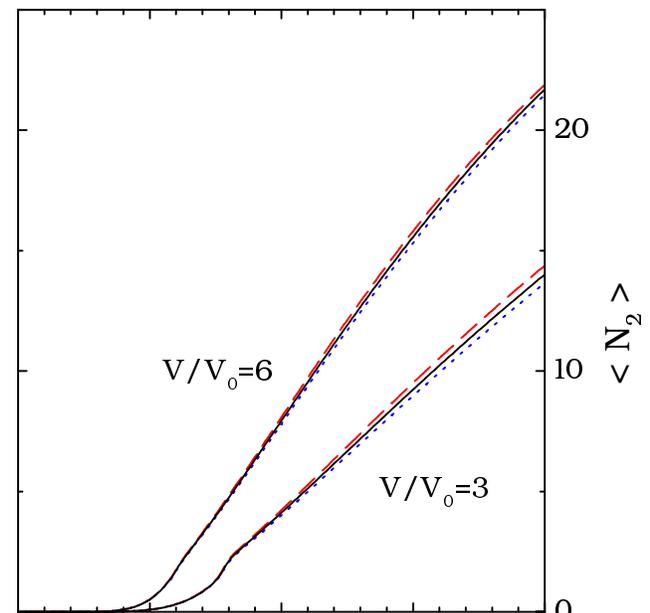
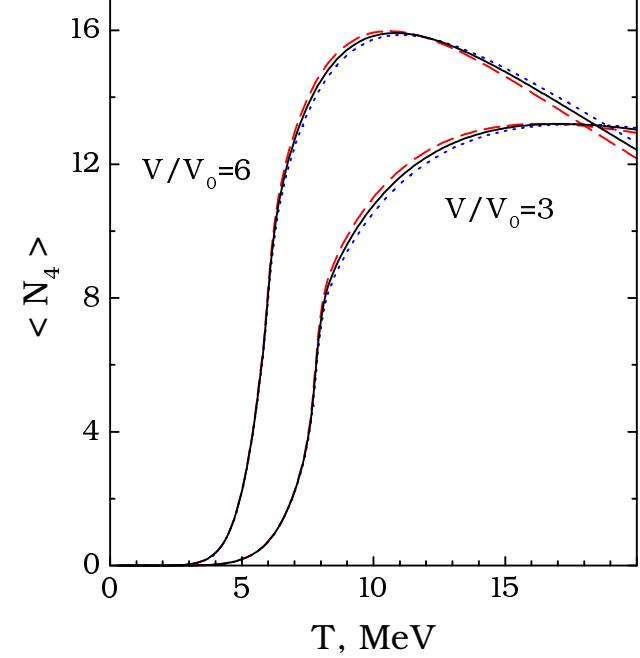
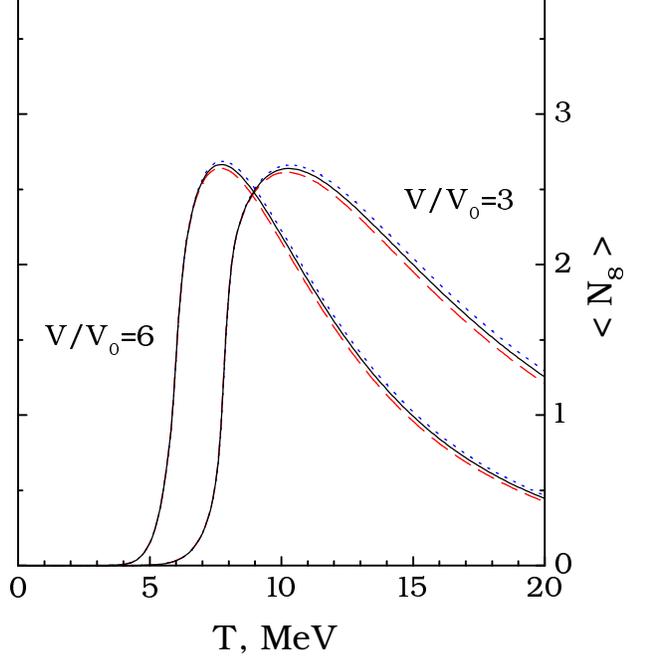
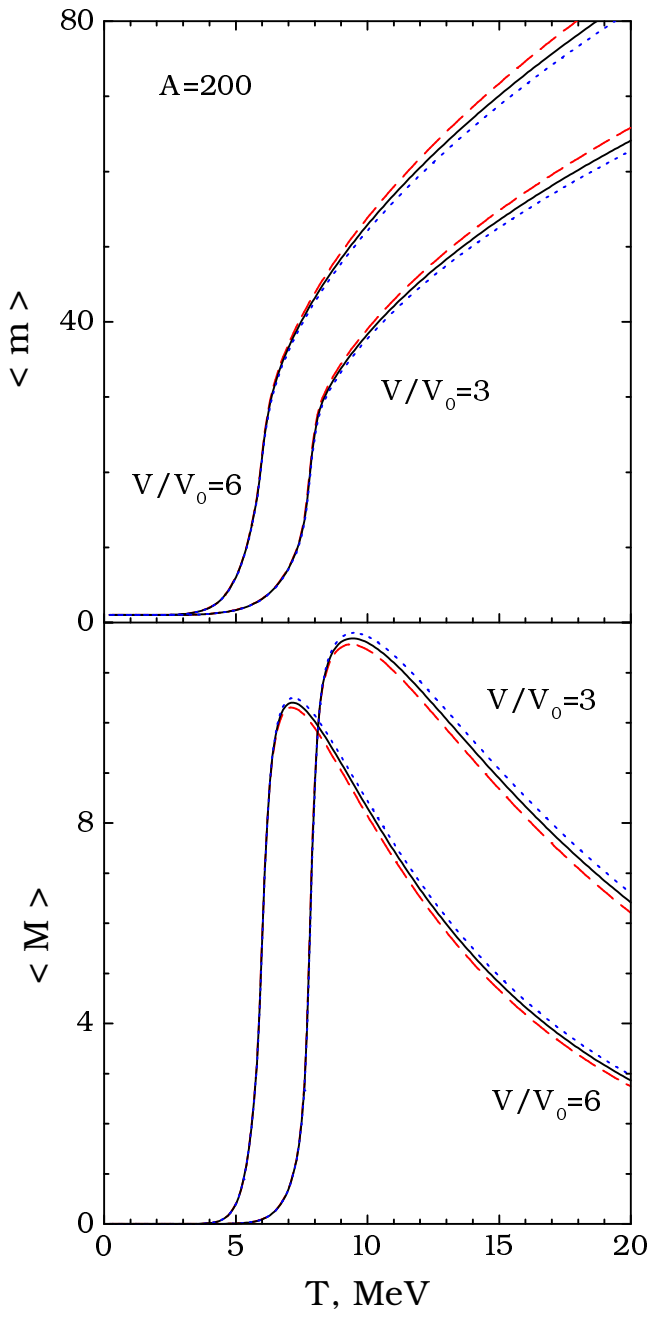
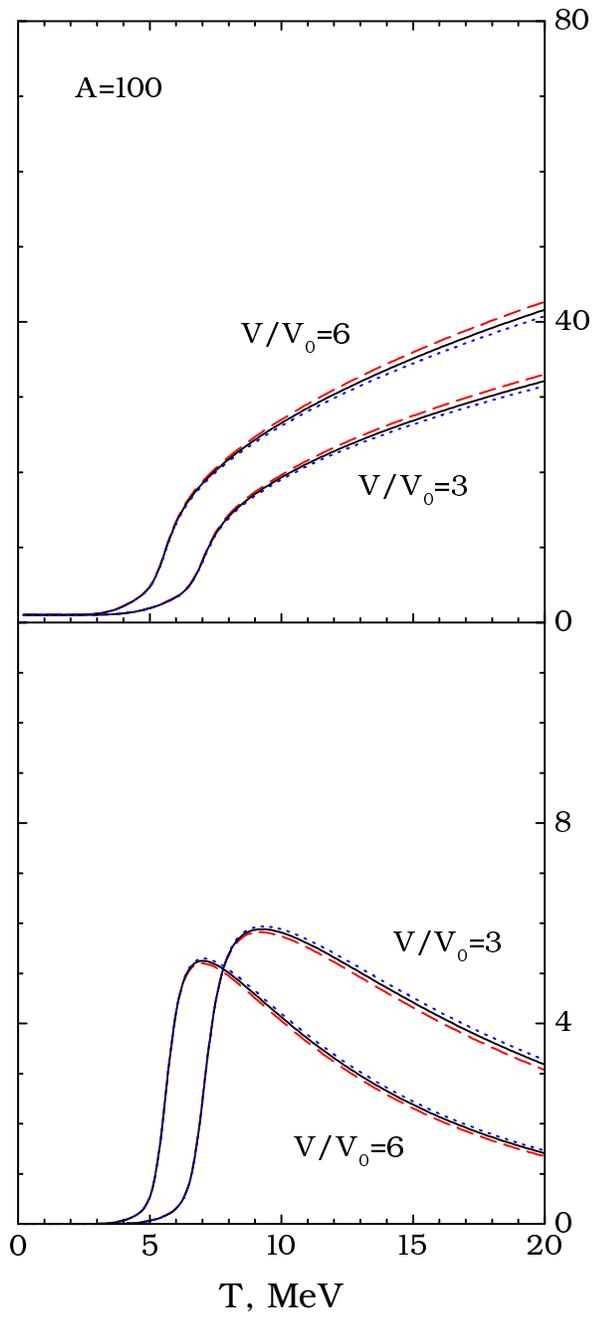
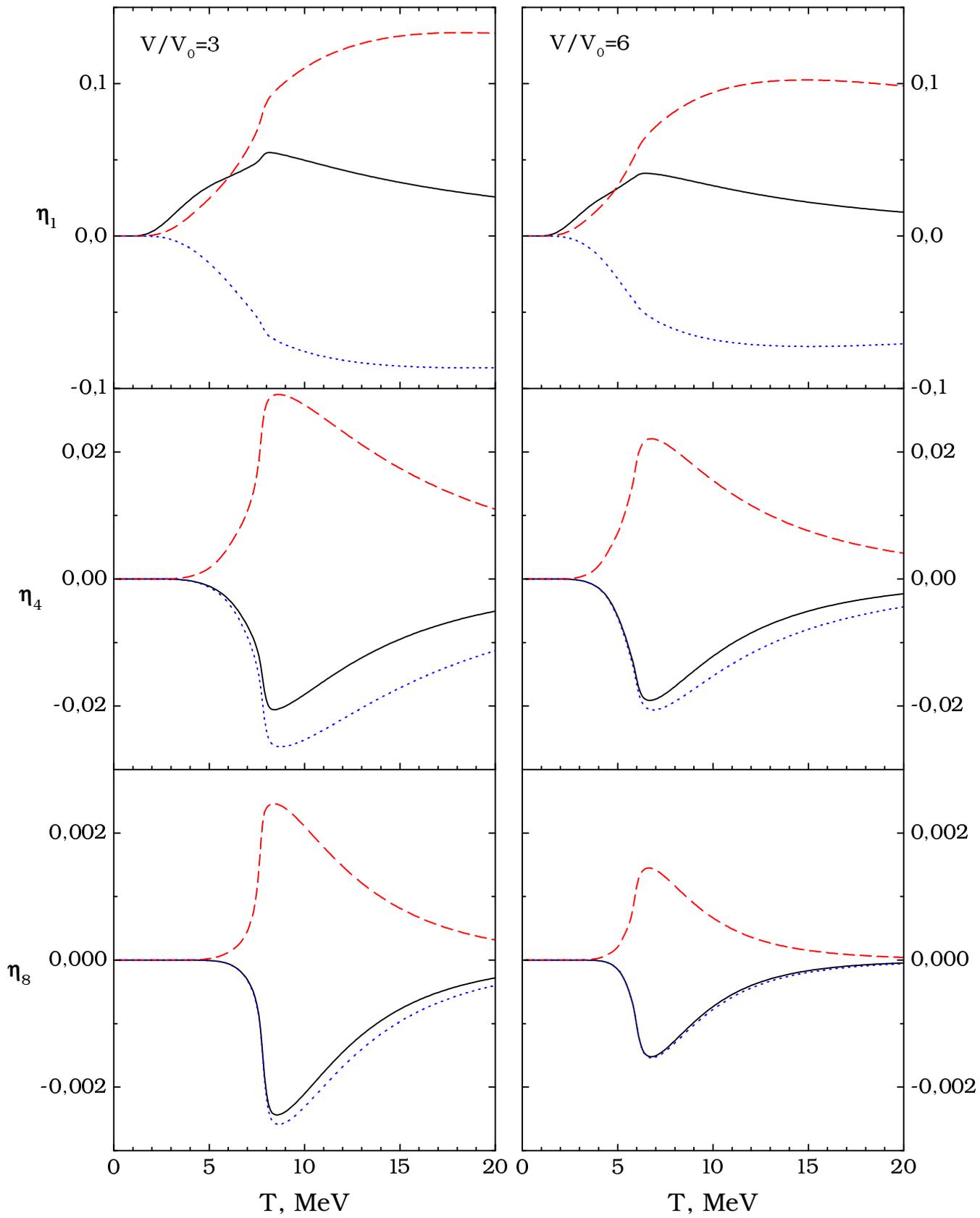
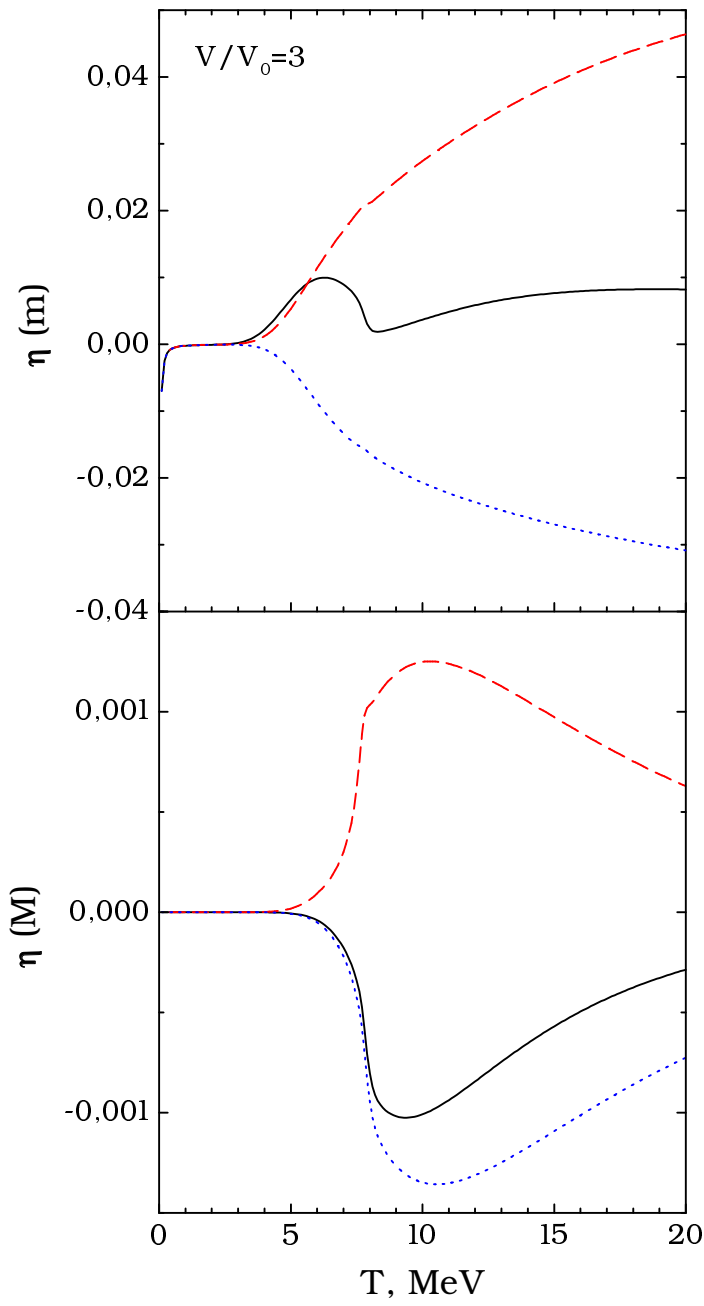
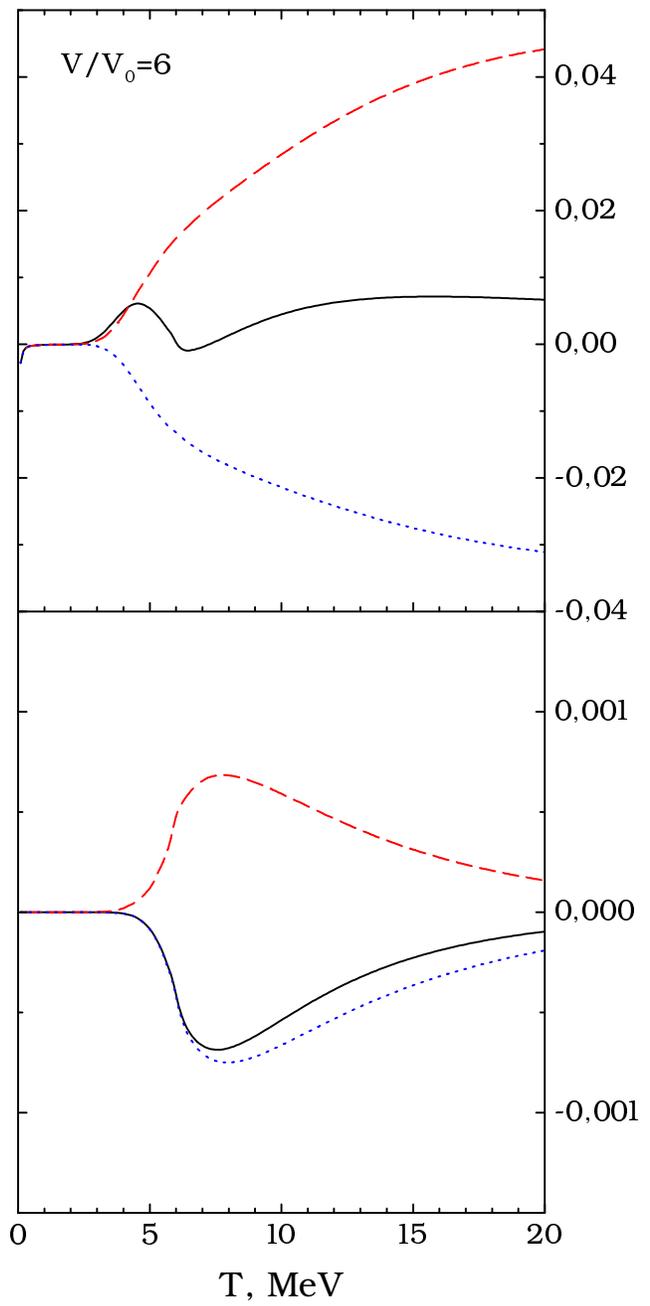
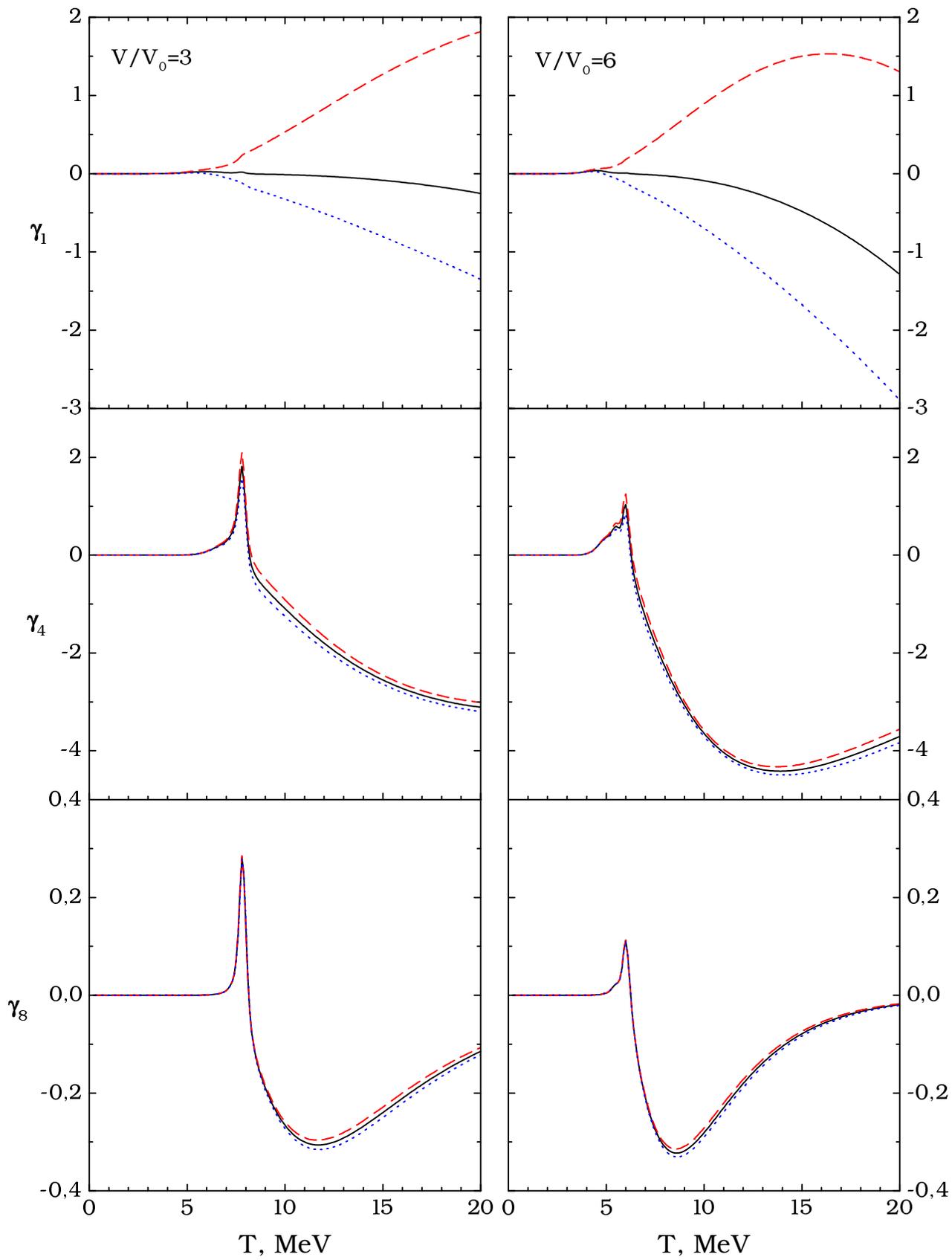
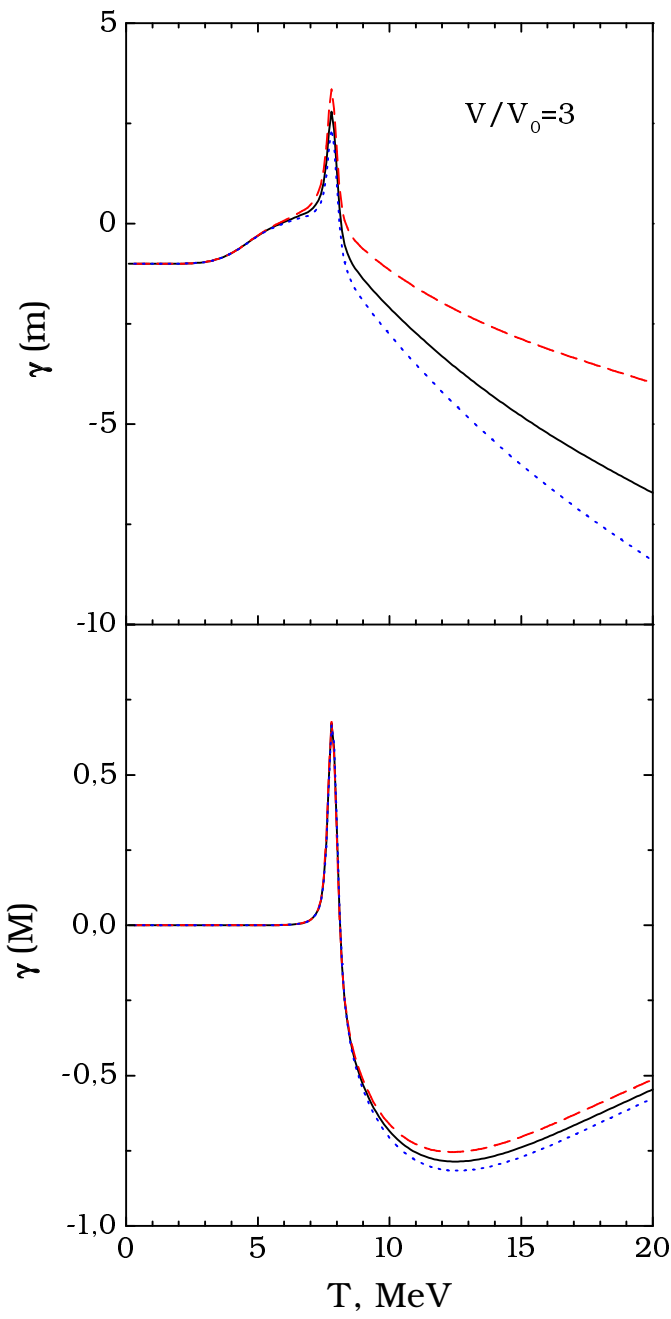
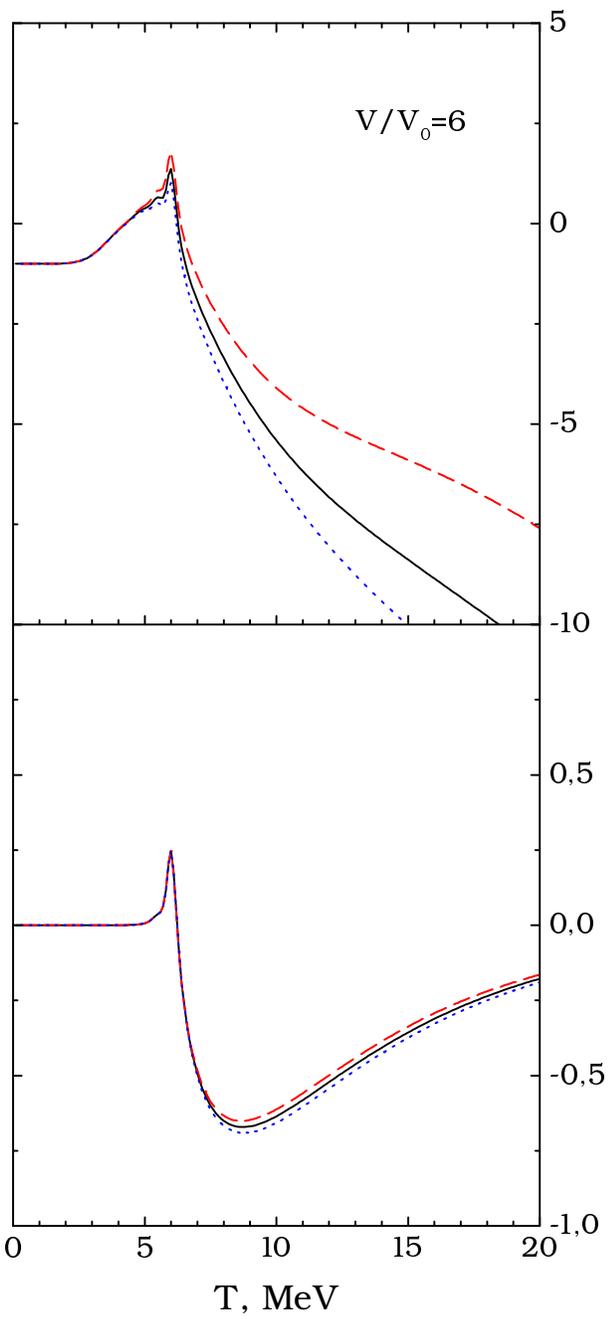
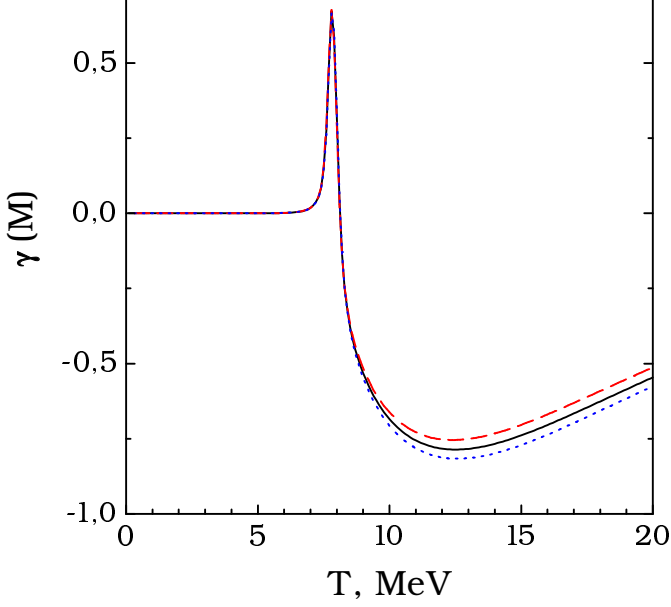
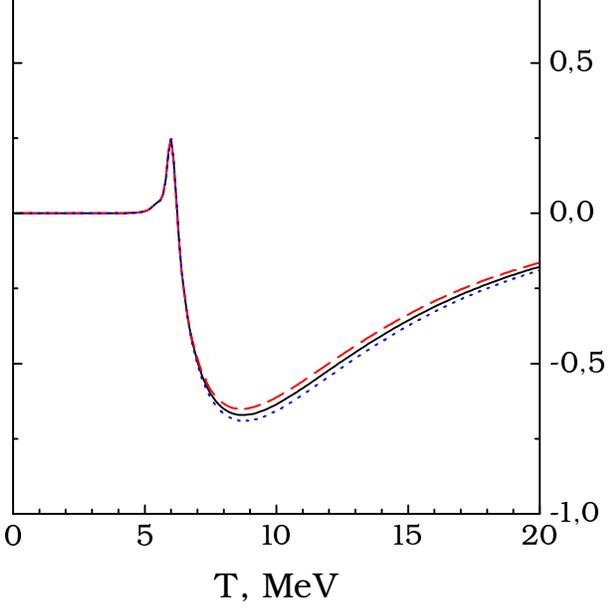
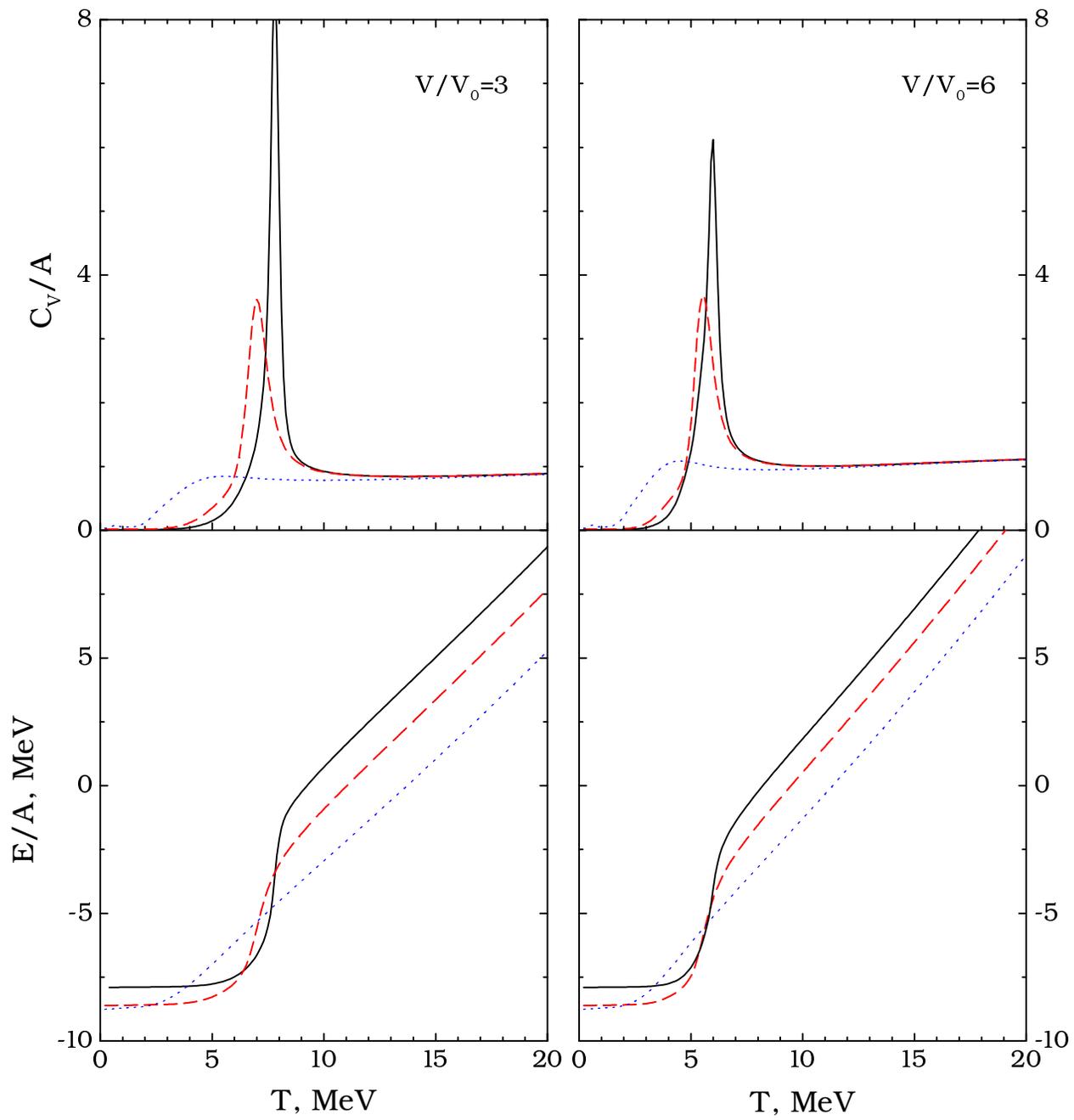
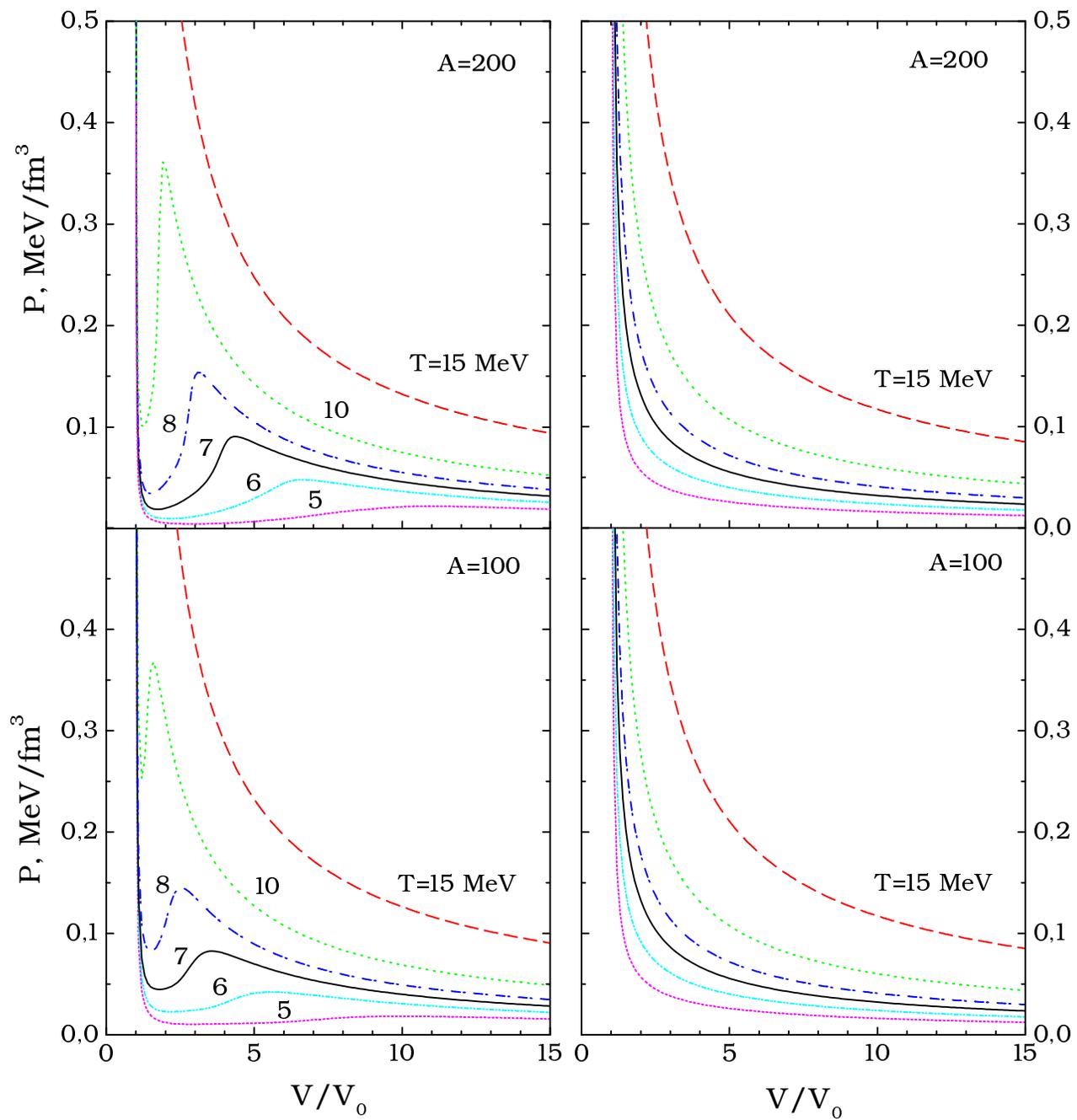
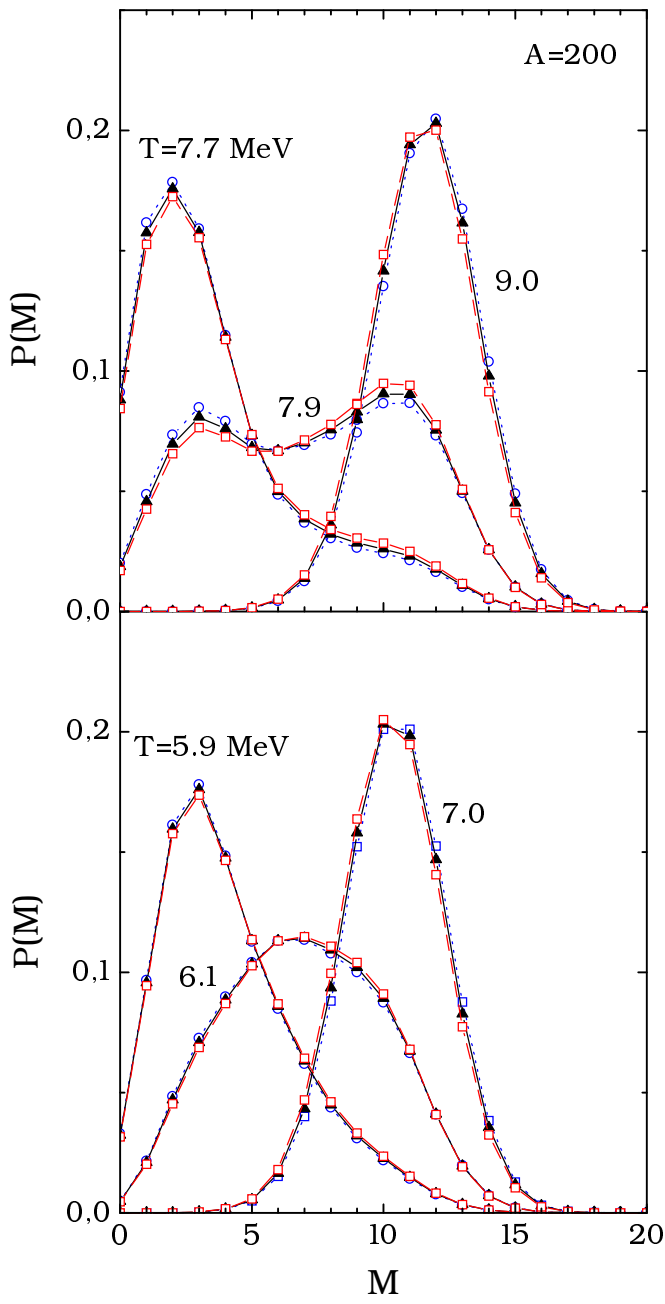
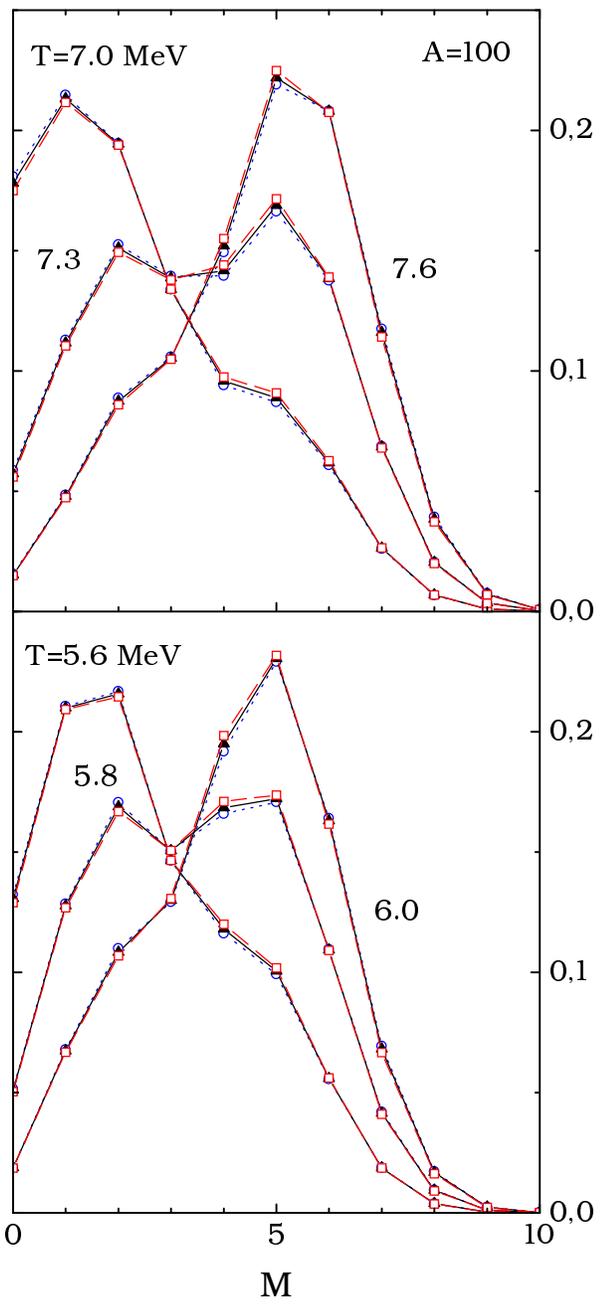
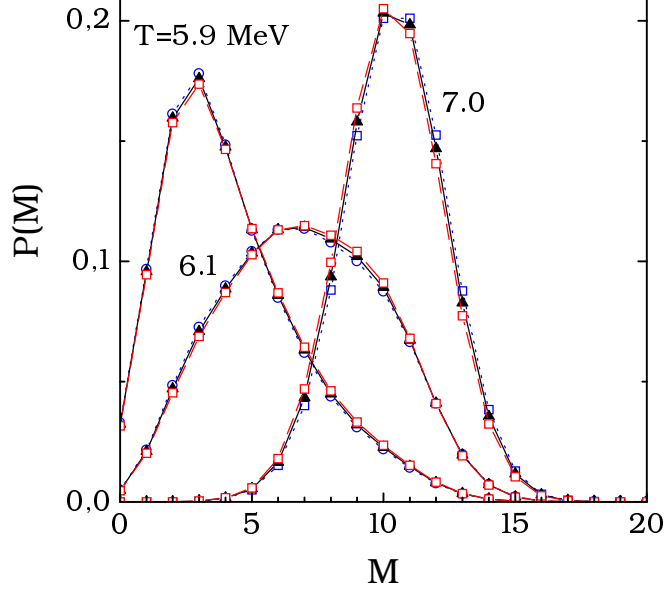
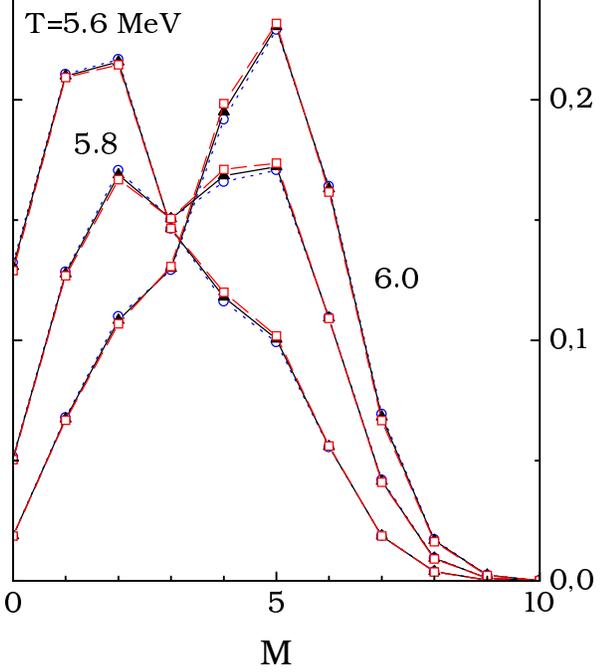
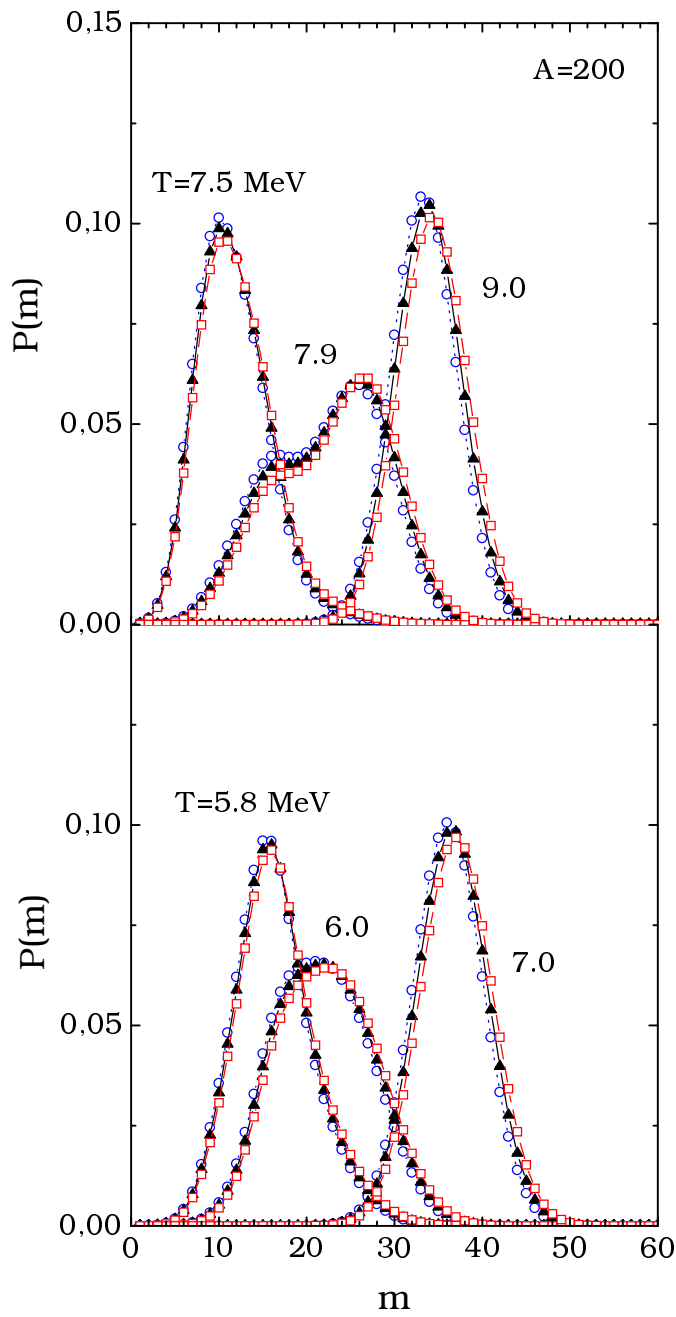
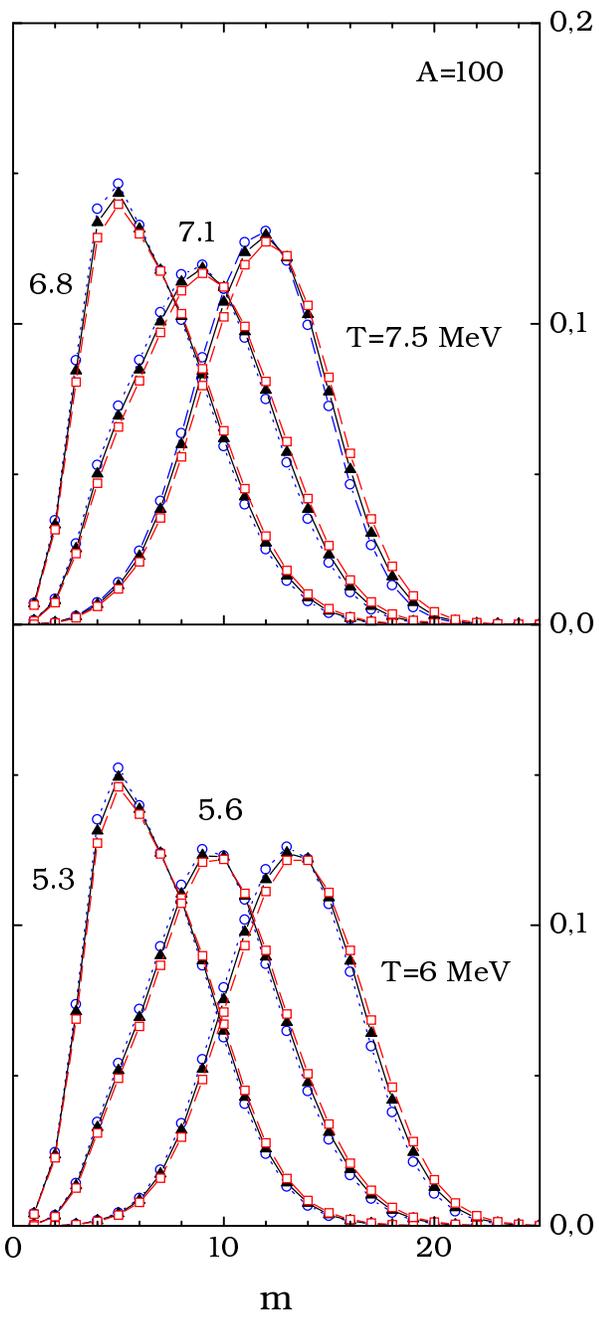
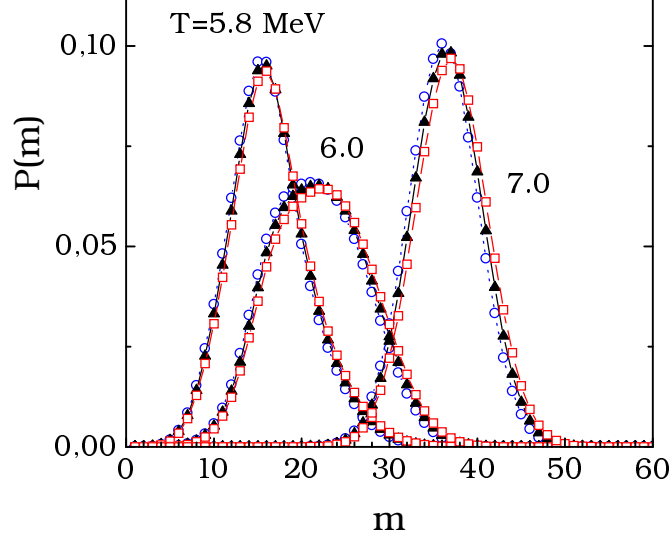
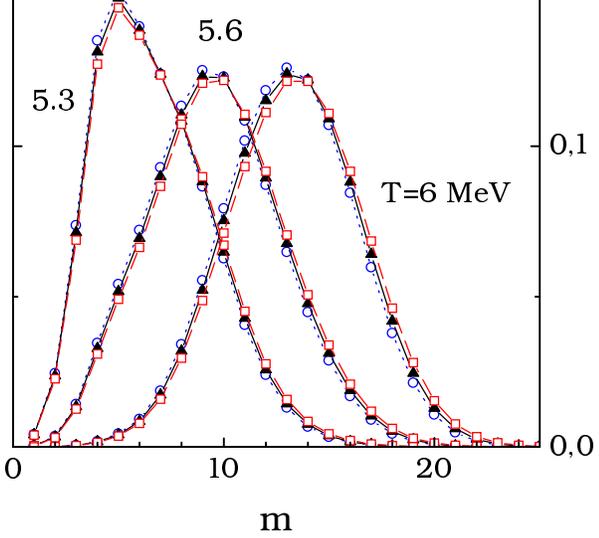
\end{document}